\documentclass[aps,prb,reprint,showpacs,superscriptaddress,groupedaddress]{revtex4-1}
\usepackage{graphicx}
\usepackage{amsmath}
\usepackage{amssymb}
\usepackage{dcolumn}
\usepackage{dsfont}
\usepackage{latexsym}
\usepackage{rotating}
\usepackage{bbm}
\usepackage[usenames,dvipsnames]{xcolor}  
\usepackage{float}
\usepackage{epsfig}
\usepackage{psfrag}
\usepackage{natbib}
\usepackage{bm}
\usepackage{eucal}
\usepackage{mathrsfs}
\usepackage{braket}
\usepackage{enumerate}
\usepackage{longtable}
\usepackage{subfigure}
\usepackage{bm}
\usepackage{hyperref}
\usepackage{amsfonts}
\usepackage{dcolumn}
\usepackage{bm}
\usepackage{subfigure}

\newcommand{\be}{\begin{equation}}
\newcommand{\ee}{\end{equation}}
\newcommand{\bn}{\begin{eqnarray}}
\newcommand{\en}{\end{eqnarray}}

\usepackage{color} 

\usepackage{hyperref}
\hypersetup{
colorlinks=true,final=true,
        linkcolor=red,
        citecolor=blue,
        filecolor=blue,
        urlcolor=blue,
}
\begin{document}

\author{S. Acharya$^{1}$}\email{swagata@phy.iitkgp.ernet.in}
\author{M. S. Laad$^{2}$}\email{mslaad@imsc.res.in}
\author{A. Taraphder$^{1,3}$}\email{arghya@phy.iitkgp.ernet.in}
\title{Manifestations of Strange Metallicity in Inelastic Neutron Studies}
\affiliation{$^{1}$Department of Physics, Indian Institute of Technology,
Kharagpur, Kharagpur 721302, India.}
\affiliation{$^{2}$Institute of Mathematical Sciences, Taramani, Chennai 600113, India
}
\affiliation{$^{3}$Centre for Theoretical Studies, Indian Institute of
Technology Kharagpur, Kharagpur 721302, India.}

\begin{abstract}
   Emergence of an orbital-selective Mott phase (OSMP) found in multi-band correlated systems leads to a non-perturbative obliteration of the Landau 
Fermi liquid in favor of a novel metallic state exhibiting anomalous infra-red (branch-cut) continuum features in one- and two-particle responses.
We use a combination of $(1)$ dynamical mean-field theory (DMFT) using the continuous-time-quantum Monte-Carlo (CTQMC) solver for a two-band Hubbard model  
and $(2)$ analytic arguments from an effective bosonized description to investigate strange metal features in inelastic neutron scattering studies for cuprates.  Specifically, 
restricting our attention to symmetry-unbroken metallic phase, we study how
emergence of an OSMP leads to qualitatively novel features in $(i)$ the dynamical spin and charge susceptibilities, and $(ii)$ phonon response in the strange metal, in detail.  
Extinction of the Landau quasiparticle pole in the one-electron propagator in the OSMP mirrors the 
emergence of critical liquid-like features in the dynamical spin response.  
This novel finding also underpins truly anomalous features in phonon dynamics, which we 
investigate by coupling half-breathing phonons in the specific context of cuprates 
to such a multi-electronic continuum.  We find good understanding of various 
anomalies encountered in experimental inelastic neutron scattering studies in the
near-optimally doped cuprates.  We also extend these results in a phenomenological way to argue how modification of phonon spectra in underdoped cuprates can be reconciled with
proposals for a nematic-plus-$d$-wave charge modulation order in the pseudogap state.  
We also study the issue of the dominant ``pair glue'' contributions to  superconductivity, 
allowing us to interpret recent pump-probe results within a strange metal scenario.  
 
\end{abstract}


\maketitle

Quantum Phase Transitions (QPT) of increasingly diverse types continue to
present a fundamental challenge to our rapidly evolving understanding of
theory of electrons in quantum matter~\cite{hertz,coleman}.  These range from the
famed itinerant type, described by the Hertz-Moriya-Millis (HMM)~\cite{hertz} model, to the local type, based on 
Heisenberg-Kondo~\cite{si} or selective-Mott~\cite{pepin,senthil} models.
Finding of the exotic strange metal phase in cuprates and certain $f$-electron systems is 
by now a fundamental problem, related to the issue of 
how strong electronic correlations can cause a basic transmutation in the nature of elementary excitations in a $d>1$
($d$ is the spatial dimensionality) metallic system, as implied by the change of the analytic 
structure of the charge- and spin-fluctuation propagators from an infra-red pole to branch-cut structure.  
Focussing on spin fluctuations, 
the following issue arises.  In a traditional Hertz-Moriya-Millis (HMM) view of QCP, 
the dynamical spin susceptibility (the Fourier 
transform of the spin-spin propagator) $\chi({\bf q},\omega)$ shows the critically overdamped paramagnon 
form for certain set of critical ${\bf Q}$s associated with magnetic order.  
In stark contrast, anomalous scale-invariant and predominantly $\omega$-dependent 
response characterizes the strange metal, where $\chi_{loc}(\omega)\simeq T^{-\eta}f(\omega/T)$ with $0<\eta<1$ is seen. 
This form appears in Kondo-RKKY models, but it is necessarily linked to proximity to magnetic order.
On the other hand, the FL$^{*}$ theory~\cite{senthil} does not {\it need} such proximity to get this form. 
Experimentally, while unconventional superconductivity (USC) in $f$-electron systems 
indeed maximises near an antiferromagnetic quantum phase transition, 
USC with high-$T_{c}$ in hole-doped cuprates sets in and
maximises far from an AF-QCP, but close to an optimal doping where a topological 
Fermi surface reconstruction (FSR) occurs.  Thus, proximity to AF-QCP 
may not be a (uniquely) necessary requirement for generating the soft 
electronic glue that results in HTSC.  This prompts a set of issues: $(i)$ what are the microscopic 
origins of the anomalous spin-fluctuation spectrum in the normal and the enigmatic pseudogap phases in cuprates, 
and, in particular, how are we to understand its specific link to equally anomalous
fermiology and transport in one picture?, $(ii)$ what non-Landau symmetry-breaking, 
if any, underlies emergence of strange metallicity and the pseudogapped metal? and $(iii)$ what are 
its consequences for HTSC that peaks precisely around optimal doping in the strange metal?

  These experimental findings and the above discussion motivate our present work.  Specifically, here we study 
 these by extending previous work on an extended periodic Anderson model~\cite{laad,civelli}.  Previous works found either a 
FL$^{*}$ metal or a local QPT between a heavy LFL with very small quasi-particle weight ($z_{FL}<<1$) and a locally critical 
metal with $z_{FL}=0$ and an anomalous $\omega/T$-scaling of the local 
propagators as the ratio $V_{fc}(k)/U_{fc}$ was varied.  
Similar behavior was also found, due to $J_{H}$, even earlier~\cite{georges}.  
This was shown to emerge as a direct consequence of effects akin to the orthogonality 
catastrophe accompanying a selective-Mott localization, a la hidden-FL~\cite{anderson} or
FL$^{*}$~\cite{senthil} theories.  The relevance of the emerging branch-cut singular propagators for understanding strange metallicity in 
cuprates has long been emphasized by Anderson~\cite{anderson} in what amounts to a kind of local approximation in the
$U=\infty$ fixed point for the Hubbard model.  In a selective-Mott metal, such 
features arise for a range of realistic parameters via strong inelastic scattering between Mott-localized and metallic subsets of the
dressed spectral function.  Here, we show how the same selective-Mottness idea also enables a natural 
understanding of the unique branch-cut continuum form of
spin fluctuations as a consequence of the dualistic (itinerant-localized) character of carriers.  

  We now describe our formulation for $\chi({\bf q},\omega)$ for the 
EPAM in detail.  Begin with the DMFT solution for the one-electron propagators
of the EPAM, defined as before~\cite{pepin,laad}
\be
H=H_{band} + H_{int} + H_{hyb}
\ee
where

$H_{band}=\sum_{k,\sigma}\epsilon_{k}c_{k,\sigma}^{\dag}c_{k,\sigma} + \sum_{<i,j>}t_{ff}f_{i,\sigma}^{\dag}f_{j,\sigma}$, $H_{int}=U_{ff}\sum_{i}n_{if\uparrow}n_{if\downarrow} + U_{cc}\sum_{i}n_{ic\uparrow}n_{ic\downarrow} + \sum_{i,\sigma,\sigma'}U_{fc\sigma\sigma'}n_{if}n_{ic}$ and $H_{hyb}=\sum_{k,\sigma}V_{fc}(k)(f_{k}^{\dag}c_{k}+h.c)$.
  
   Here $\epsilon_{k},t_{ff}$ are kinetic energies, $U_{cc}, U_{ff}, U_{fc}$ are local Hubbard interactions (intra and inter orbital) and
$V_{fc}$ is inter-orbital hybridization terms for $c$ and $f$ electrons respectively. 
In the $f$-electron context, this is an extended periodic Anderson model (EPAM) which 
incorporates the important effect of $f$-valence fluctuations.  
In the context of high-T$_{c}$ cuprates, models very similar to Eq.(1) have a long history: 
originally proposed~\cite{varma} in the context of marginal-FL theory, they also emerge from ab-initio 
quantum chemical (QC) calculations~\cite{liviu}, as effective models for nodal and anti-nodal states~\cite{imada}, 
and in the context of a proposal for the hidden order in the famed pseudogap phase of underdoped 
cuprates~\cite{giamarchi}.  Especially interesting is the observation that the two-band 
Hubbard model in QC calculations closely resembles Eq.(1) if we 
relabel $c_{k,\sigma}\rightarrow d_{x^{2}-y^{2},\sigma}$ and $f_{i,\sigma}\rightarrow d_{3z^{2}-r^{2},\sigma}$ 
(the small $f-f$ hopping corresponds to the much smaller direct hopping between the $d_{3z^{2}-r^{2}}$ 
states compared to that between $d_{x^{2}-y^{2}}$ states, 
and the renormalized hybridization is found to be sizable, 
and to have a $d$-wave form factor).  Interestingly, this is also
precisely the model related to the one used by Weber {\it et al.}~\cite{giamarchi} in the 
context of loop-current order being a primary candidate for the hidden order in the pseudogap phase of cuprates.
Thus, all results and discussions below with model (1) in mind can be applied to cuprates in the light of the above. 

   Application of our results to cuprates is thus based on: $(i)$ applicability of such a two-band model, with
$c=d_{x^{2}-y^{2}}$ and $f=d_{3z^{2}-r^{2}}$ rests upon earlier ab-initio results~\cite{liviu}, 
where exactly such an effective model was extracted from a quantum chemical calculation, 
$(ii)$ applicability of single-site DMFT, at least for the strange metal without the complication of 
having either short-ranged spatial correlations (as in cluster-DMFT studies) or 
broken symmetry, both of which will be relevant at lower $T$.
Finally, in recognition of the fact that strange metal anomalies are seen in optimally doped cuprates, 
we start out with partially filled $f,c$ bands (see below for precise numbers). 
The quasi-local criticality view is implicit in both, 
hidden-Fermi liquid~\cite{anderson} and marginal-Fermi liquid~\cite{cmv} views
for the cuprates.  Thus, our DMFT approach needs extension in the pseudogap (PG) phase in underdoped cuprates.  
In this work, we adopt a phenomenological approach to study the PG phase, and a full theoretical study of this
phase awaits extension of multi-band-DMFT to include spatial correlations. 
This is a demanding task, and is left for future studies.

   Here, with these comments in mind, we investigate the strange metal phase, focussing on dynamical one- and two-particle responses.  
Our main results are: $(i)$ for sizable $U(U')$, we find that both exhibit a critical continuum with infra-red 
singular behavior up to high energy, characteristic of the strange metal, $(ii)$ we 
qualitatively describe resistivity, optical conductivity and magnetic fluctuation data 
in near-optimally doped cuprates without further assumptions, and $(iii)$ we 
build upon these positive features to study phonon responses in detail, providing a novel interpretation of 
phonon anomalies in INS data in terms of coupling of bond-stretching phonon modes to a 
critical continuum of multi-particle excitations.  
Further, motivated by recent pump-probe studies, we also use these
results to study the issue of relative importance of such critical electronic vis-a-vis phononic 
glues for pairing in cuprates.  We conclude by discussing how novel local quantum criticality in the 
OSMP phase is drastically different from famed Hertz-Moriya-Millis (HMM) views that obtain 
close to quantum phase transitions to quasi-classical ($e.g$, antiferromagnetic) order.
  
\vspace{0.0em}
\begin{figure}[h!]
\centering
\subfigure[]{\label{f:C11}\epsfig{file=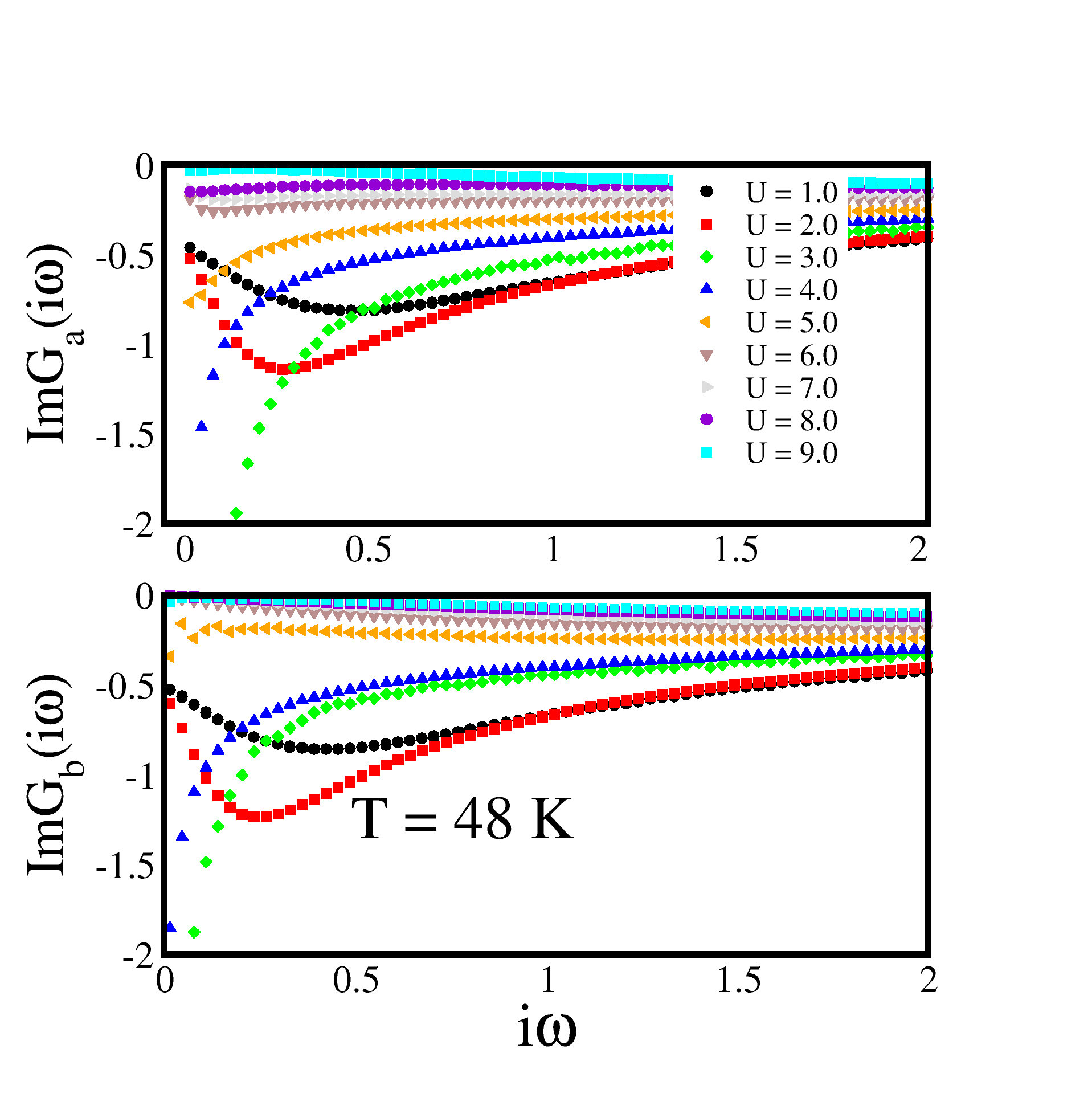,trim=0in 0in 0in 0.0in,
clip=true,width=0.980\linewidth}}\hspace{-0.0\linewidth}
\subfigure[]{\label{f:C21}\epsfig{file=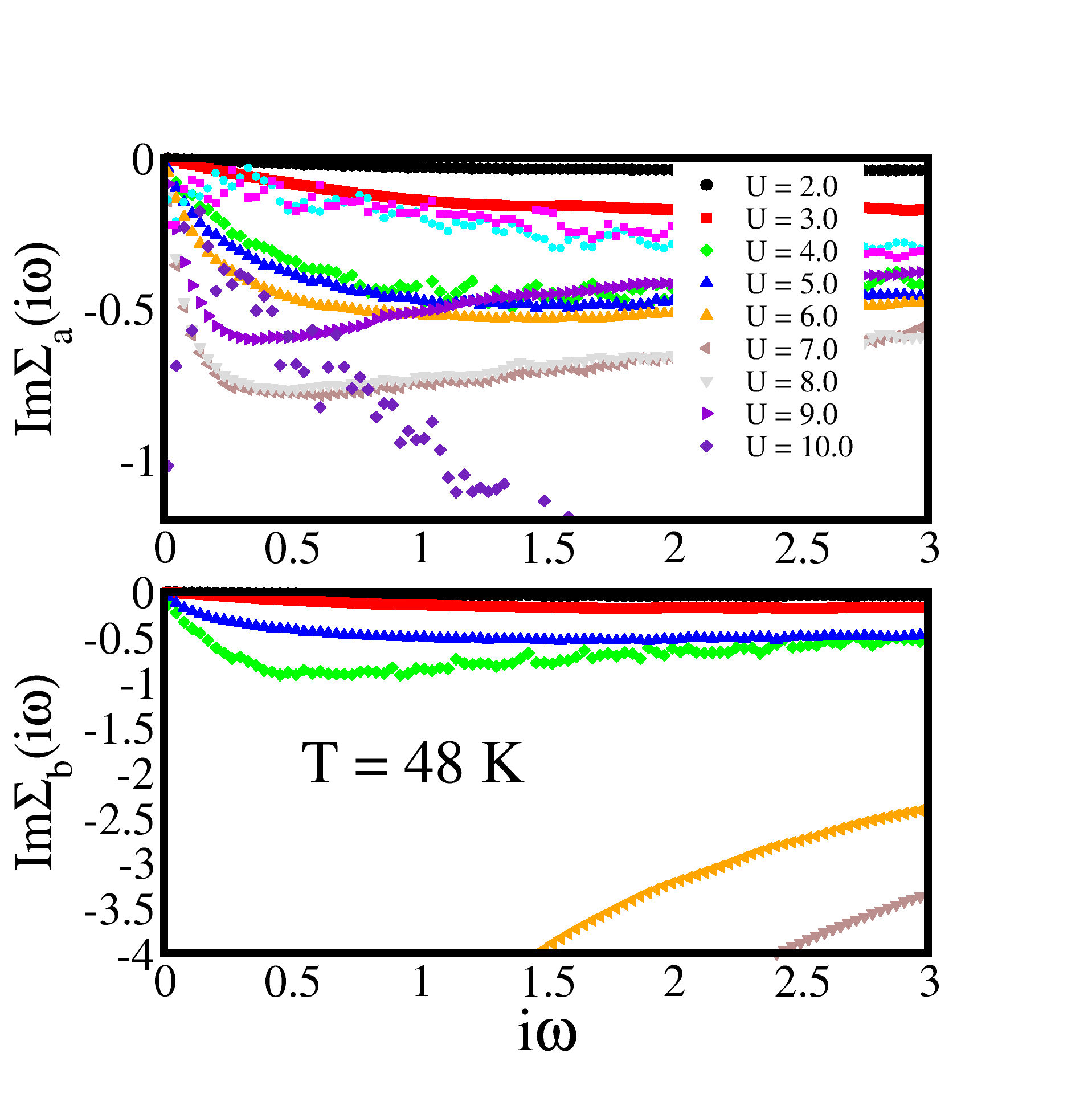,trim=0in 0in 0in 0.0in,
clip=true,width=0.980\linewidth}}\hspace{-0.0\linewidth}
\caption{$ImG_{a,b}$$(\omega)$ and Im$\Sigma_{a,b}(i\omega_{n})$ as a function of
$i\omega$ for a two-orbital square lattice with inter orbital hybridization $t_{ab}$ (witha d-wave form factor)
are shown as a function of $U$.}
\label{fig1}
\end{figure}

  It is natural to work in the rotated $a,b$ fermion basis (linear combinations of $c,f$ fermionic operators) 
which diagonalizes the non-interacting part of $H$ above).  
For the free band-structure, the $d$-wave form of $V_{fc}(k)$ generates a pseudogapped 
one-electron density-of-states (DOS) (not shown). The occupancies of the $a,b$ bands are 
found to be $0.52,1.48$ at $U$ and $U' (= U_{ab}/3) = 0$.
For $U_{fc}<U_{fc}^{(c)}$ a stable strongly correlated LFL phase 
was found in earlier work using iterated perturbation theory (IPT) as an impurity solver in DMFT.  
However, for $U_{fc}>U_{fc}^{(c)}$, a selective-Mott metal with an infra-red 
branch-cut form of the one-electron propagator, i.e, a local quantum critical 
metal, emerged as a consequence of the strong scattering induced OC. 
Given the approximate nature of the IPT solver, it is obviously of interest to 
ask whether these appealing features survive upon use of a more exact impurity solver.
Here, we have solved $H$ (Eq.1) within DMFT using the much more exact continuous-time quantum 
Monte Carlo (CT-QMC) solver.  The advantage here is that the local component of both, one- and two-particle dynamical responses can be reliably computed, in contrast to IPT where two-particle 
responses (susceptibilities) need reliable knowledge of the fully dynamical but local irreducible vertex.  
This is presently a demanding task, unless they can be argued to be irrelevant, as in large-$N$
approaches to DMFT~\cite{sachdev-ye}.

\vspace{0.0em}
\begin{figure}[h!]
\centering
\subfigure[]{\label{f:C11}\epsfig{file=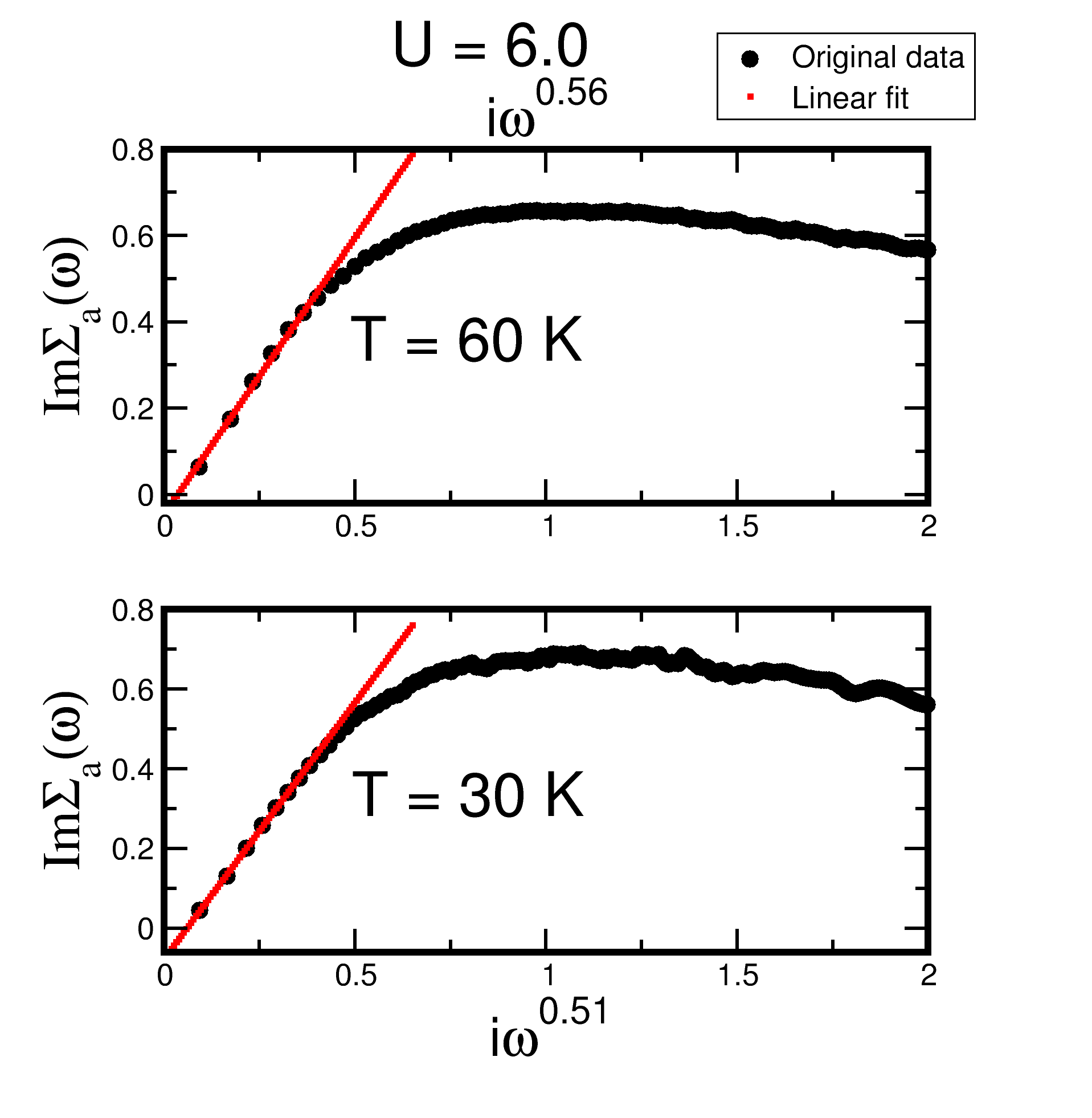,trim=0in 0in 0in 0.0in,
clip=true,width=0.98\linewidth}}\hspace{-0.0\linewidth}
\caption{$Im\Sigma_{a}(i\omega)$ as a function of $i\omega$ for a two-orbital 
square lattice with inter orbital hybridization $t_{ab}$ and $U=6.0$ at low $T$.  
Clear power-law behavior in Im$\Sigma_{a}(i\omega)=C(i\omega)^{1-\eta}$ with $\eta=0.44, 0.49$ 
for $T = 60 K, 30 K$ up to high energy $O(0.5)$~eV, is visible.  This is a specific characterization of strange metallicity.}
\label{fig2}
\end{figure}

  In the present work, we obtain the full vertex-corrected two-particle dynamical 
responses for the EPAM using the hybridization expansion-based CTQMC method as an impurity solver~\cite{alps}.
This part is based on our earlier study of an OSMP in a generalized two-orbital Hubbard model~\cite{jpcx}. 
  Here, we explicitly show that the spin and charge susceptibilities possess a long-time (in imaginary-time) 
quantum-critical form, and use a trick inspired by conformal 
symmetry to extract an effective analytic form in real frequencies. 
Such a trick has been widely employed in other contexts~\cite{senthil}. 
This will allow us to discuss transport and magnetic fluctuations in the strange metal phase, 
and to discuss underlying microscopic features leading to the specific anomalies, in detail.  

  In Fig.~\ref{fig1}, we show the imaginary parts of the one-fermion Green functions for the 
$a,b$ orbital states, along with the corresponding self-energies. 
It is clear thereby that as $U$ and $(U'=U_{ab}=U/3)$ increase, 
the semimetallic behavior for $U=0$ (due to the ${\bf k}$-dependent hybridization) 
changes into a normal, albeit correlated LFL metallicity at 
moderate $U$. 
Beyond $U=4.0$, however, an orbital-selective differentiation of electronic states clearly reveals itself 
in both Im$G_{a,b}(i\omega_{n})$ and Im$\Sigma_{a,b}(i\omega_{n})$: the $a$-states are in an 
incoherent metal state, while the $b$-states are in the Mott insulating regime. 
This dualistic feature exists in the whole parameter region $4.0<U<9.0$ for the given set of 
one-electron parameters in the EPAM, and thus this is an orbital-selective Mott phase, rather than a QC point. 
For $U>10.0$, we find that a full Mott insulator phase obtains.  Interestingly, the $a,b$ orbital occupancies
change continuously as $U$ and $U'$ are raised: at $U=3.0,U'=1.0$, we find $n_{a}=0.82, n_{b}=1.18$.  This trend persists throughout the correlated LFL metal regime ($U\leq 4.0$), beyond which ($U=5.0$) we find that 
$n_{a}=1=n_{b}$.  Remarkably, this is a curious manifestation of emergent particle-hole symmetry in the OSMP.  It directly rationalizes the onset of selective-Mott physics: the half-filled narrower band ($b$ states) selectively undergoes a Mott transition, while the broader ($a$-band) band states remain bad-metallic.
It is interesting that there are tantalizing hints of such p-h symmetry around optimal doping from 
$dc$ Hall data in cuprates~\cite{ong}.  Within one-band Hubbard model contexts, this has been accounted 
for by using extended one-electron hopping integrals beyond nearest neighbors, 
and realizing restoration of p-h symmetry at a critical doping (where the Fermi energy sits at the 
saddle-point at $K=(\pi,0)$ in the bare one-electron dispersion).  
Our multiband model has inbuilt p-h asymmetry, and it is very interesting that restoration of p-h symmetry 
in the above sense goes hand-in-hand with OSMP and emergence of strange metallic behavior.
 
\vspace{0.1em}
\begin{figure}[h!]
\centering
\subfigure[]{\label{f:C11}\epsfig{file=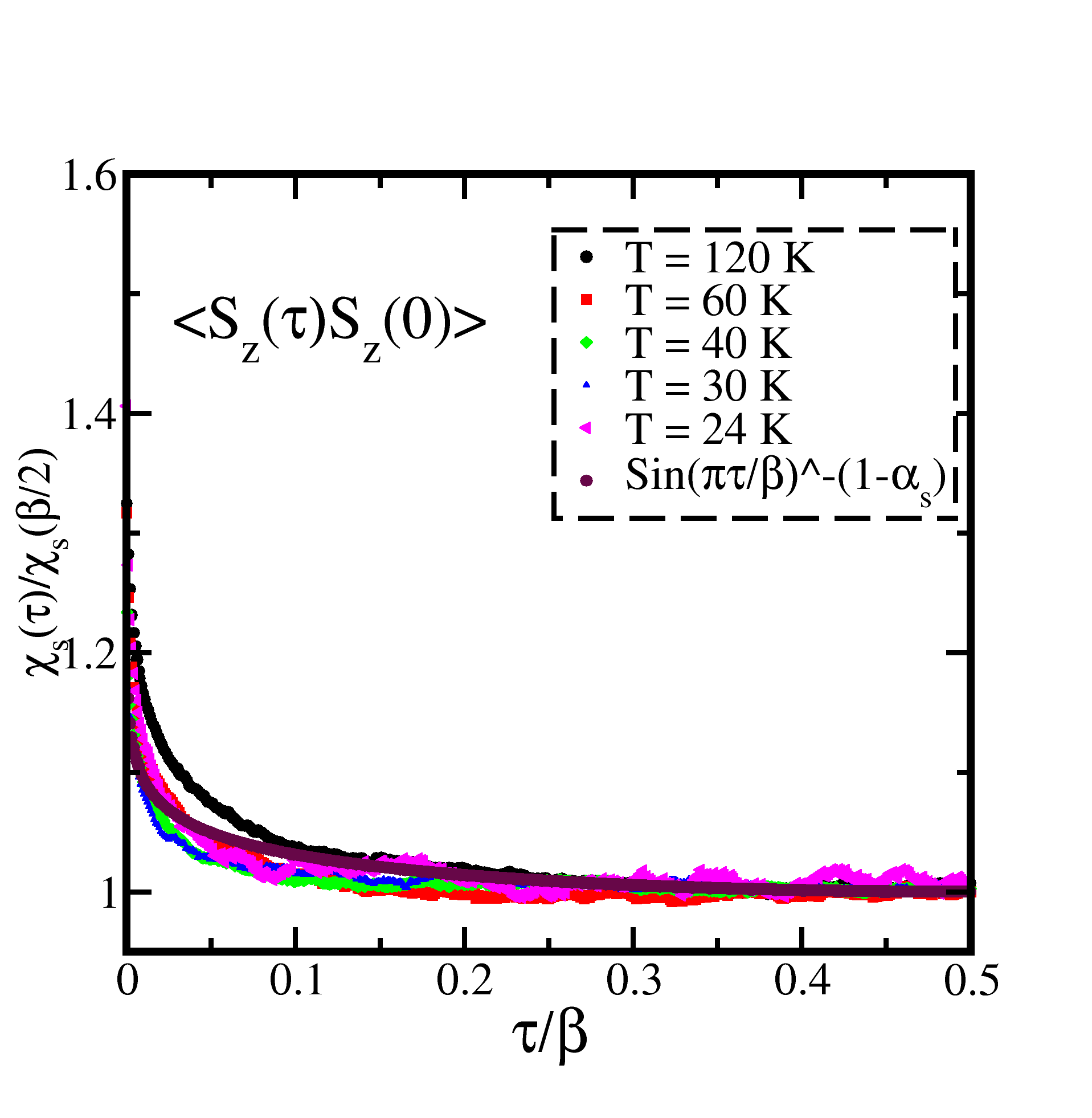,trim=0in 0in 0in 0.0in,
clip=true,width=0.98\linewidth}}\hspace{-0.0\linewidth}
\subfigure[]{\label{f:C21}\epsfig{file=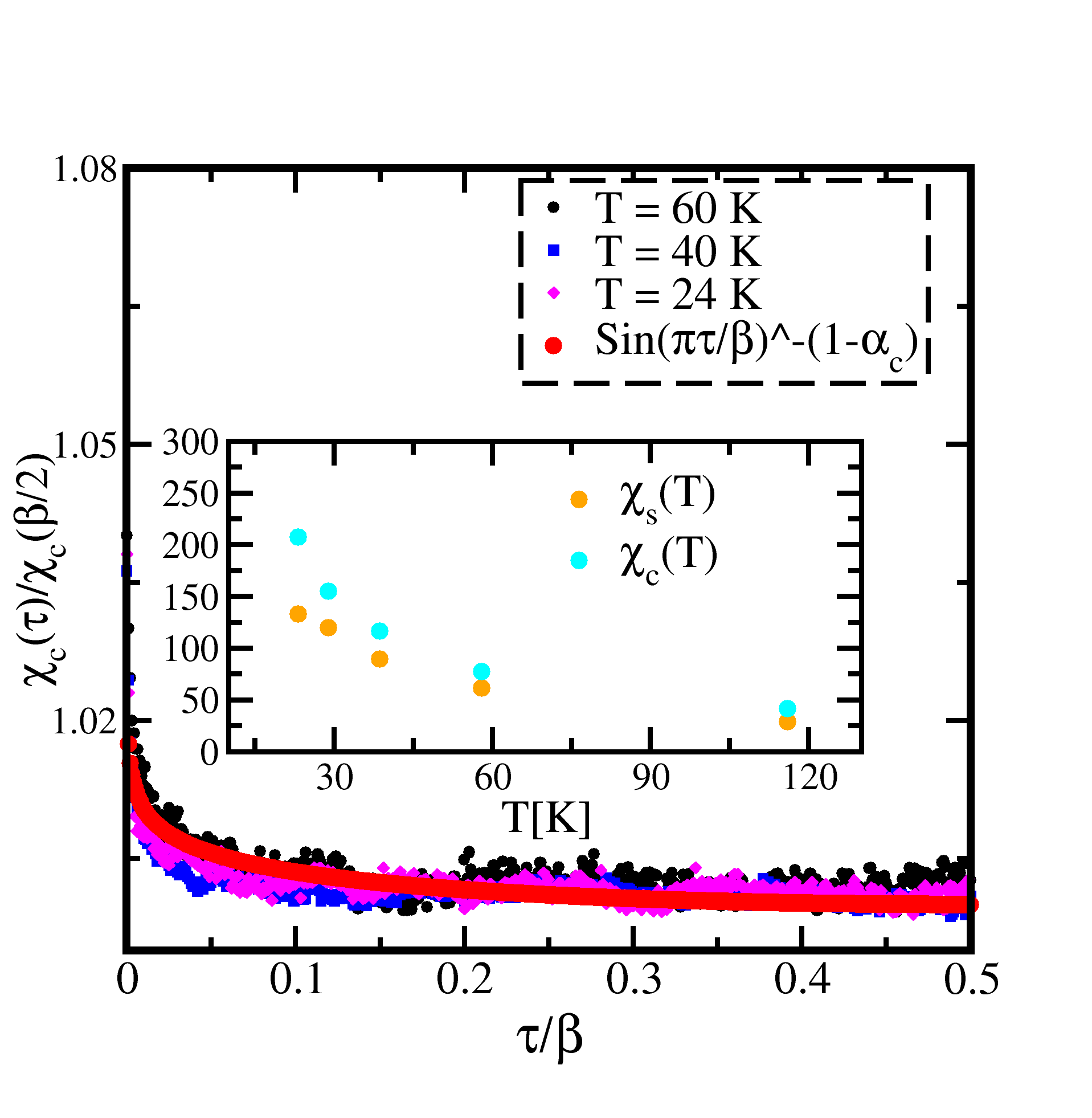,trim=0in 0in 0in 0.0in,
clip=true,width=0.98\linewidth}}\hspace{-0.0\linewidth}
\caption{$Im\chi_{s,c}(\tau)$ as a function of $\tau/\beta$. 
Both, the dynamical spin (s) and charge (c) susceptibilities show 
quantum critical scaling, with $\chi_{s,c}(\tau)/\chi_{s,c}(\beta/2) \simeq [$sin$(\pi\tau/\beta)]^-{(1-\alpha_{s,c})}$,
with $\alpha_{s}(0.9)\neq \alpha_{c}(0.98)$.
Within our numerical accuracy, this constitutes a high-$D$ realization of ``spin-charge separation''.
The inset shows the static local
susceptibilities where $\chi_{c}(T)$ is more singular than $\chi_{s}(T)$ at low temperatures.}
\label{fig3}
\end{figure}

  Focussing on the OSMP, we now exhibit the $a,b$-fermion self-energy, 
Im$\Sigma_{a,b}(i\omega_{n})$, and the full
local spin susceptibility, Im$\chi_{s}(i\omega_{n})$, 
for the orbital-selective Mott phase Fig.~\ref{fig1}.~\ref{fig2} and Fig.~\ref{fig3}. 
Interestingly, we find clear evidence of anomalous fractional power-law exponents in both. 
In Fig.~\ref{fig2}, we see
that $Im\Sigma_{a}(i\omega_{n})\simeq (i\omega_{n})^{1-\eta}$, 
with $\eta=0.44 (T = 60 K)$ and $\eta=0.49 (T = 30 K)$. 
In fact, this behavior extends up to rather high energies $O(0.5)$~eV, which is 
characteristic of a multi-particle electronic continuum expected to extend to high energies in a quantum critical regime.  
Remarkably, the local part
of the dynamical spin- and charge susceptibilities also exhibit infra-red singular 
and fractional power-law scaling behavior, characteristic
of the strange metal. We see this as follows. 
Using the DMFT(CTQMC) results, 
we see that all spin susceptibility ($\chi_{s}(\tau)$) curves for $U>4.0$ 
almost fall on each other and, upon a careful fitting with a scaling function 
characteristic for quantum criticality, 
we find that $\chi_{s}(\tau)/\chi_{s}(\beta/2)$ scales as $[$sin$(\pi\tau/\beta)]^{-(1-\alpha_{s})}$ 
with $\alpha_{s}=0.9$ Fig.~\ref{fig3}. This is precisely the scaling form dictated by conformal 
invariance in the quantum critical region.  It exactly corresponds to the scaling form

\be
\chi_{s}''(\omega) \simeq \omega^{-\alpha_{s}}F_{s}(\omega/T)
\ee
with $F_{s}(x)=x^{\alpha_{s}}|\Gamma(\frac{1-\eta}{2}+i\frac{x}{2\pi})|^{2}$sinh$(x/2)$, 
whence we immediately read off that $T^{-\alpha_{s}}\chi_{s}''(\omega)$ is a universal scaling 
function of $\omega/T$, with a fractional-power-law dependence.  
Such a unique form for the dynamical spin susceptibility near magnetic QCPs is experimentally 
seen in more than a few systems by now, specifically in (not obviously proximate to magnetic order), 
CeCu$_{6-x}$Au$_{x}$~\cite{schroder}, among others.  In cuprates, a similar form has long been known
from early studies of dynamical spin fluctuations~\cite{aeppli}, 
where $\alpha_{s}\simeq 1$, in good qualitative accord with our extracted exponent.  
On a theoretical level, extinction of Landau fermionic quasiparticles in $G_{b}(k,\omega)$ 
in the OSMP directly manifests in the emergence of a critical branch-cut continuum in 
(one-spin-flip) spin-fluctuations. 
Even more interestingly, the dynamical charge susceptibility also 
exhibits similar scaling form, but with an 
exponent $\alpha_{c} \neq \alpha_{s}$, implying that spin and charge fluctuations 
propagate with distinct velocities.  This is an explicit realization of a high-dimensional spin-charge separation.
This is our central result, and we show below that it leads to very good accord with 
the unusual normal-state magnetic fluctuation spectra and transport in 
strange metals. 
Physically, these features emerge as a direct manifestation of emergent, critical pseudoparticles 
driving the extinction of stable Landau-damped FL-like collective modes  in the strange metal. 
This is simply because charge and spin fluctuations are themselves constructed from the now 
incoherent multiparticle continuum, rather than usual Landau quasiparticles.  
\vspace{0.1cm}

  What is the underlying physical origin of these emergent anomalous features?
  As in FL$^{*}$ or Kondo breakdown theories, emergence of critical liquid-like features do not 
need a one-to-one association with proximity to $T=0$ magnetic order, 
since DMFT only accesses strong local dynamical correlations. 
In the OSMP, $V_{fc}(k)$ plays an especially distinct role.
    Explicitly, the tendency of $V_{fc}$ to transfer a $a$-fermion into an $b$-fermion 
band is dynamically blocked in the OSMT. 
This is because the lower-Hubbard band now 
corresponds to all singly occupied $b$-states, so action of $V_{fc}$ must create a doubly occupied  
(two opposite-spin electrons in the $b$-orbital) intermediate state. 
However, this lies in the upper
Hubbard band in the $b$-sector and thus the resulting term now has the form 
$V_{fc}'(n_{i,b,-\sigma}b_{i,\sigma}^{\dag}a_{j,\sigma}+h.c)$, which couples the $a$-fermion to a Gutzwiller-projected $b$-fermion,
and thus has no interpretation in terms of a coherent one-electron-like state any more. 
In this sector, this is a high-energy state, and is asymptotically projected out from the low-energy Hilbert space 
(space of states with energy less than the selective-Mott gap). 
It is this emergent projective aspect that is at the root of irrelevance of 
$V_{fc}(k)$ at one-electron level 
(thus, this is a Kondo destruction~\cite{pepin}) and emergence of strange metal features we find above. 
Thus, we find a mechanism for the destruction of the LFL picture to be very closely related to the 
hidden-FL view of Anderson~\cite{pwa}. 
In absence of a lattice Kondo scale, the hopping of an $b$-electron from site $i$ to the bath or vice-versa 
(and similarly for the $a$-electron) thus
creates a suddenly switched-on local potential for the $b$-electrons.  
A direct upshot is that the inter-band
spin fluctuations (now local triplet excitons of the type $(a_{i\sigma}^{\dag}b_{i\sigma'}+h.c)$ with $\sigma'=-\sigma$) 
thus also experience a local potential that is (incoherently) switched on and off as a function of time. 
The resulting problem is precisely the inverse of the Anderson-Nozieres-de Dominicis
X-ray edge problem in the local limit of DMFT, and is a central feature of Anderson's hidden-FL theory~\cite{pwa}.
At two-particle level, this reflects itself in a divergent number of soft, local spin fluctuation modes, manifesting itself 
as an infra-red singularity along with local quantum critical $\omega/T$ scaling and 
anomalous exponents in the spin fluctuation spectrum. 
This is completely borne out by our CTQMC results (Fig.~\ref{fig3})  
  The resulting spin polarizability, 
$\Pi(\omega) \simeq \chi_{zz}^{-1}(\omega) \simeq A|\omega|^{1-\alpha}$ in the QC metal,
in stark contrast to $\Pi({\bf q},\omega) \simeq -ia\omega$, known in the heavy
 LFL metal.
The fully renormalized $\chi({\bf q},\omega)$ now reads

\be
\chi({\bf q},\omega)=\frac{1}{\Pi(\omega)+J({\bf q})}
\ee
where $J({\bf q})\simeq -c+({\bf q}-{\bf Q})^{2}+...$ (e.g, $Q=(\pi,\pi)$ for Neel AFM).  At finite $T$,  $\Pi(\omega,T)=AT^{\alpha}f(\omega/T)$.

\vspace{0.4cm}

{\bf Physical Responses In The Strange Metal}

\vspace{0.4cm}

Remarkably, it turns out that our results afford a consistent and surprisingly good quantitative description 
of both transport and neutron results in the strange metal.  
Let us consider the consequences of our numerical findings in more detail in this section, 
with a specific focus on cuprates.

\noindent $(i)$  The dynamical spin susceptibility shows an explicit $\omega/T$-scaling with anomalous fractional exponent $\alpha_{s}$.  It is qualitatively consistent with observations in near optimally doped cuprates~\cite{aeppli},
and may also be applicable to other $f$-electron systems where there is independent evidence of the relevance of strong, quasilocal liquid-like critical fluctuations near destruction of magnetic order.  From the form of the dynamical spin susceptibility, we can also immediately read off that the spin relaxation rate in nuclear-magnetic resonance (NMR) studies will vary very weakly with $T$, as $1/T_{1}\simeq T^{0.1}$ or that 
$1/T_{1}T \simeq T^{-0.9}$.  This is quite close to the measured $1/T_{1}T \simeq T^{-1}$ in near-optimally
doped cuprates~\cite{aeppli}.

\noindent $(i)$  The fact that Im$\Sigma_{a,b}(i\omega)\simeq (i\omega)^{1-\eta}$ with $(1-\eta)=0.51,0.56$ for $T = 30 K, 60 K$ allows study of $dc$ and $ac$ conductivities without any further approximation within DMFT. 
This is because irreducible vertex corrections to the current correlation functions for the conductivities in the Bethe-Salpeter equation rigorously vanish in this limit.
Transport is now entirely determined by the simple bubble diagram composed of the full DMFT one-electron propagators. 
Explicitly, the $dc$ resistivity is now $\rho_{dc}(T)\simeq T^{2(1-\eta)}=T^{1.02}$ for 
$T = 30 K$ and $T^{1.12}$ for $T = 60 K$.  Such an unusual resistivity, $\simeq T$, has long been known as one
of the defining features of the strange metal in cuprates.  The optical conductivity can also be readily estimated as $\sigma(\omega)\simeq \omega^{-2(1-\eta)}=\omega^{-1.02}, \omega^{-1.12}$. 
This is indeed consistent with data in underdoped cuprates as well as in the $D=3$ 122-Fe arsenides~\cite{akrap}.  Interestingly, Akrap {\it et al.} associate this with novel self-energy effects for nodal quasiparticles in UD cuprates.  In our modelling, the nodal-versus-antinodal aspect enters via the ${\bf k}$-dependent $f-c$ hybridization.  In our model, this would correspond to an inter-site hybridization between $d_{x^{2}-y^{2}}$ and $d_{3z^{2}-r^{2}}$ orbitals having a $d$-wave form factor.  Interestingly, this is precisely the form extracted from ab-initio quantum chemical calculations, which forms the basis for our two-band modelling for cuprates (see earlier discussion in context of our choice of $H$). 
OSMP are known to be generic in Fe arsenides by now~\cite{capone}, and transport anisotropy~\cite{ourFeAs} as well as strain and angle-resolved photoemission (ARPES)~\cite{fisher} studies show clear signatures of localization of $d_{xz}$ orbital states in UD Fe-arsenides.   A two-band model such as one we use, but with different hopping matrix elements and a different ${\bf k}$-space form factor for the non-local $xz-yz$ hybridization could exhibit similar features in its OSMP phase.  Our work now explicitly links the anomalous optical response, at least in 122-Fe arsenides, to such an underlying OSMP. 

  Such results have also been discussed earlier in the context of Kondo-RKKY models and in the FL${*}$ context.  
However, our findings are also somewhat different from earlier works.

  $(i)$ The above constitutes an important difference from the 
Kondo-RKKY case, where the power-law fluctuational form is tied down to $D=2$.  
In our case, this is not at all necessary, since the excitonic 
singularity results from the OC that 
accompanies selective-Mottness in DMFT, which can happen in $D=3$ as well.  

  $(ii)$ Moreover, though in  both cases, the physical reason behind emergence of the 
critical metal involves critical Kondo destruction, 
there is still an important difference: in the Kondo-RKKY case, it is linked 
to proximity to AF order as a fall-out of the Kondo-RKKY competition.  Here, it is caused by the lattice OC due to selective Mott criticality and not 
necessarily due to proximity to AF 
order.  

  $(iii)$ Thus, such behavior can be observed near AF order as a
fall-out of selective Mottness, but need not always be so: it could also arise close to a $T=0$ valence instability (${\bf Q}=0$): $e.g$, in $\beta$-YbAl$_{4}$.
In fact, strong $U_{fc}$ which is necessary to get the local strange metal
also implies strong quantum fluctuations of the $f$-valence. 
In the latter case, the appropriate charge (valence) susceptibility will 
diverge in exactly the same way.  At finite-$T$, we have

\be
\Pi(\omega,T)=AT^{\gamma}g(\omega/T)
\ee
with $\gamma=(1-\alpha)$ and $g(x)$ is a universal scaling function of $x$.

\vspace{0.5cm}

{\bf Analytical Insight into DMFT Results}

\vspace{0.5cm}

  This leads to an insight, first formalized by Anderson~\cite{anderson,casey} that we now use to proceed further.  We refer interested readers to Casey and Anderson's papers for relevant details used in this section.
  
Since the local impurity problem can be bosonized in radial channels
on a $(1+1)D$ half-line~\cite{schotte}, 
$\chi({\bf q},\omega)\simeq \omega^{-\gamma}f(\omega/T)$ can also be 
understood, using the Schotte {\it et al} trick~\cite{schotte}, 
as arising from a set of radial spin density tomonagons 
(collective bosonic modes composed of spin- and charge density modes) 
emanating from every local site in an effective equivalent 
high-D view of the fermionic system.  In fact, these can be identified with 
the critical bosons.  Referring back to Schotte 
{\it et al.}, one sees, thanks to the commutation relations 
$[H,b_{q}^{\dag}]=\omega(q)b_{q}^{\dag}$ in the low-energy subspace spanned by bosons $b,b^{\dag}$ defined below, that the
latter must in reality represent critical p-h spin fluctuation modes: in terms of the $a,b$ 
fermions, one has $b_{i}^{s,c}=(1/\sqrt{N})\sum_{k,k',a,b=f,c}b_{k,\sigma}^{\dag}a_{k',\sigma'}e^{i(k-k').R_{i}}$, with $\sigma=\sigma', \sigma'=-\sigma$ for charge (c) and spin (s) respectively: these are precisely the 
spin- and charge-density (excitonic)
variables whose singular fluctuations arise from the lattice OC in a strange metal.  Can we use this 
representation for such novel spin fluctuations to get more physical insight?

\begin{figure}[h!]
\centering
\includegraphics[angle=0,width=0.96\columnwidth]{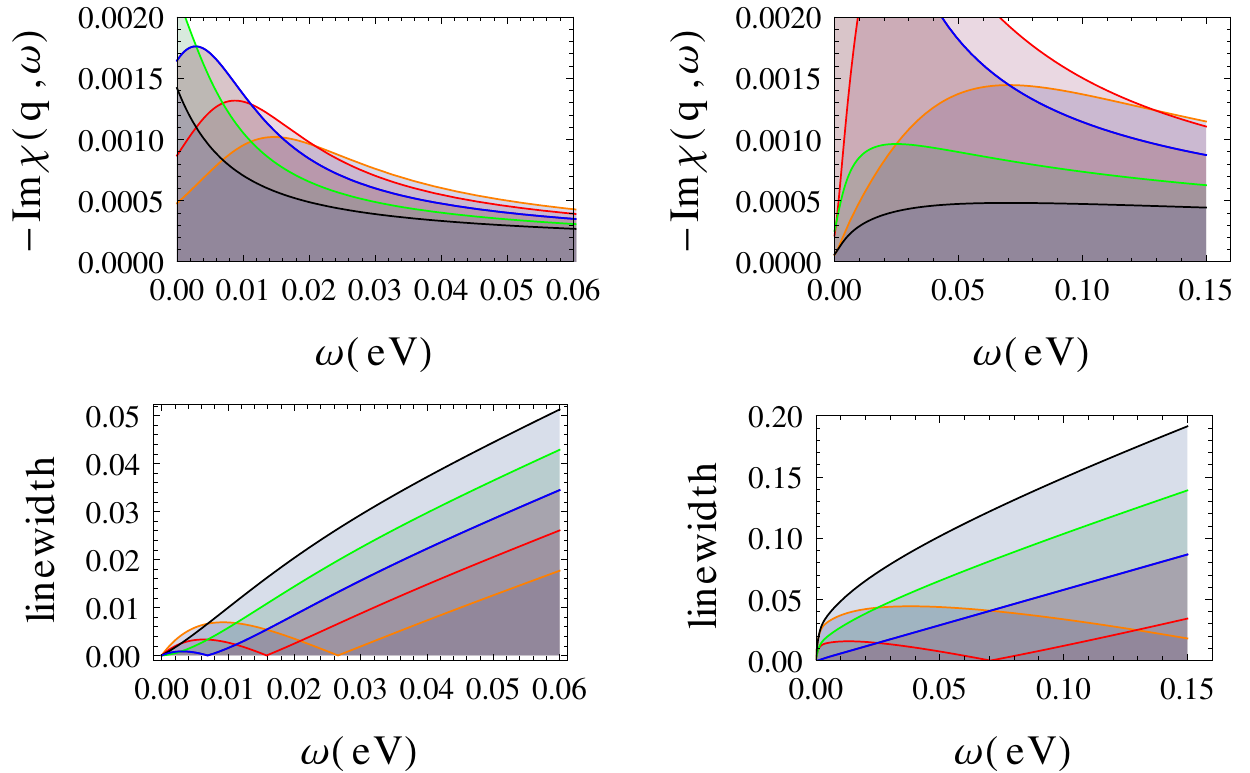}
\caption{The dynamical spin susceptibility and spin-fluctuation linewidths for the strange metal phase of 
cuprates at various ${\bf q}$ in the Brillouin zone \textcolor{orange}{(0,0)}\textcolor{red}{($\pi/2,0$)}
\textcolor{purple}{($\pi,0$)}\textcolor{green}{($\pi,\pi/2$)}\textcolor{black}{($\pi,\pi$)}\textcolor{blue}{($\pi/2,\pi/2$)} 
of the reduced Brillouin zone at $\beta$ = 100 (left panel) and 1000 (right panel).}
\label{fig4}
\end{figure}

  We now find it illuminating to use a slick trick~\cite{schotte} to show how a
related analytic procedure 
can be explicitly carried out in an especially fruitful way in the selective 
metal.  In this case, wipe-out of heavy-FL screening by the lattice 
OC is implied by the irrelevant hybridization, $V_{fc}(k)$ as discussed before. 
  Let us consider how this physics reflects itself in the tomonagon variables.  Using the definition of the tomonagons as above, we can write~\cite{anderson,casey}

\be
H'=\sum_{q}\omega(q)b_{q}^{\dag}b_{q} + (1/\sqrt{L})\sum_{q}g(q)(b_{q}+b_{-q}^{\dag})
\ee
where, crucially, the new aspect that enters in the OSMP is that 
$g(q)=g(q,t)=g(q)\theta(t)$ is now suddenly switched on 
as a function of $t$ only in the case where $V_{fc}(k)$ is irrelevant, i.e, 
precisely in the selective-Mott state.  This is because only in this case 
do the fermionic spin fluctuation modes in $H'$ see a fluctuating  
field that suddenly switches on and off as a function of $t$ (this is 
the two-particle analogue of the sudden switching-on by $U_{fc}$ in the limit 
of irrelevant $V_{fc}$ in $H$).  In bosonic lore, $H_{res}$ describes 
spin fluctuation processes: in a scattering process involving two 
magnons, the potential seen by a magnon ($b_{i}^{\dag}$) is 
time-dependent and strongly fluctuating in the manner of a step function 
because the $b$-fermion occupation switches suddenly between $n_{a}=0,1$ as a 
function of time in the OSMP.  The origin of this is that, given an irrelevant $V_{fc}(k)$,
 an $b$-electron can only be transferred into the bath (making $n_{b}=0$, 
switched off potential) or to the local site ($n_{b}=1$, potential 
switched on).  Via the definition 
$b_{i}^{\dag}=(1/\sqrt{N})\sum_{k,k'}a_{k,\uparrow}^{\dag}b_{k',\downarrow}e^{i(k-k').R_{i}}$, and the structure of $H'$, it then follows that this potential is switched on and off in a sudden way as a function of 
$t$.  The second part of $H'$ thus describes dynamic spin 
fluctuations above a quiescent filled sea of tomonagons~\cite{anderson}; we 
emphasize that 
this is, after all, a suitable low-energy representation for spin fluctuations.  But $H'$ can now be diagonalized by a~\cite{schotte} 
generalized Lee-Low-Pines (LLP) shift transformation, yielding the spectral 
function 
$g(t)=\langle T[S_{i}^{+}(t)S_{i}^{-}(0)]\rangle =$exp$[\int_{0}^{\Omega}\frac{\lambda^{2}(\omega)N(\omega)}{\omega^{2}}(e^{i\omega t}-1)d\omega]$.  Here, we 
have introduced a tomonagon DOS, $N(\omega)$ and $\Omega\simeq$ max$J(q)$, 
and interpreted $\lambda^{2}(\omega)N(\omega)$ as a 
spectral function of the tomonagons.  Interestingly, following 
Hopfield~\cite{hopfield}, the choice $\lambda^{2}(\omega)N(\omega)\simeq (1-\alpha)\omega$ reproduces the IR singularity in $\chi_{ii}^{+-}(\omega)$.  
At finite $T$, an explicit calculation gives 
$g(t) \simeq [\frac{sinh(t\pi/\beta)}{t\pi/\beta}]^{-(1-\alpha)}$, 
implying that $\chi_{ii}^{+-}(\omega)$ will show $\omega/T$-scaling.  
In particular, it turns out that at $T=0$, we recover precisely 
$\chi_{loc}(\omega) \simeq \theta(\omega)|\omega|^{-(1-\alpha)}$.  
In Fig.~\ref{fig4}, we show $\chi({\bf q},\omega)$ for various ${\bf q}$ as a 
function of $\omega$ at fixed $T$ and as a function of $T$ for
fixed ${\bf q}={\bf Q}=(\pi,\pi)$ with $J({\bf q})$ obtained from the same 
parameter choice as in earlier EPAM work.  In close correspondence with key 
findings in the normal state spin fluctuation spectrum measured in INS data for strange metals,
we observe: $(i)$ anomalously broad continuum response resulting from a 
branch-cut, rather than any conventional pole-like analytic structure of 
$\chi(q,\omega)$.  Moreover, incoherent magnetic spectral weight piles up at 
low energy in a way that respects $\omega/T$-scaling
with Im$\chi({\bf q},\omega,T)\simeq A({\bf q})T^{-(1-\alpha)}F(\omega/T)$  Simultaneously, the ${\bf q}$-dependence of the INS lineshape is entirely 
that of the unperturbed Lindhardt susceptibility of the two-band model, 
since the spin self-energy, $\Pi(\omega)$ is purely local,
$(ii)$ rigorously vanishing magnon residue, namely, $z_{m}\simeq (1-\partial_{\omega}\Pi(\omega)|_{\omega\rightarrow 0})^{-1}=0$, following from 
$\Pi(\omega)\simeq -a|\omega|^{1-\alpha}$ with $\alpha >0$, and $(iii)$ a linewidth
 that nevertheless decays linearly or sub-liearly with ${\omega,T}$ as a function of ${\bf q}$ if we define the spin fluctuation rate, 
$\Gamma_{q}(\omega)=\frac{\omega \chi'(q,\omega)}{\chi''(q,\omega)}=\pi(1-\alpha)\omega$.  Thus, the slope of the spin fluctuation scattering lifetime is simply related to the power-law exponent in $\chi_{ii}^{+-}(\omega)$, a feature which can be tested in INS work.

\vspace{0.5em}

\vspace{0.5em}

  On the other hand, when $V_{fc}(k)$ is relevant, the local Kondo screening 
must produce a heavy-FL state at low $T$: this corresponds to the quantum 
paramagnetic heavy-FL solution found in DMFT for the EPAM for 
$U_{fc}<U_{fc}^{(1)}$ in earlier work.  In this case, it is known that one must seek an infra-red solution to $H'$ with an 
adiabatically slowly switched on $g(q,t)$ instead of a suddenly switched one, 
leading, as required, to heavy-FL metallicity with small $z_{FL}$~\cite{mh,laad}.  This is because, when $V_{fc}(k)$ is relevant, an $a$-electron can now hop coherently from the site $i$ into the bath and back.  We emphasize that this route to restoration of 
one-electron Fermi liquid coherence is intimately tied down to the fact that coherent one-electron hybridization ($V_{fc}$)  in the EPAM now involves both, real charge and spin fluctuations, in contrast to the classical Kondo scenario, where $b$-charge fluctuations are suppressed in the infra-red.  This manifests 
itself in a finite recoil of the $b$-fermion during scattering, giving it a 
heavy but finite mass and cutting off the 
infra-red singularity below a low-energy scale $E_{recoil}\simeq k_{B}T_{coh}$
in both $\rho_{a}(\omega),\rho_{b}(\omega)$ in the associated impurity 
model, and this is faithfully reflected in our CTQMC results (where a correlated LFL metal obtains for $U \leq 4.0$~eV, as well as in earlier DMFT work~\cite{laad}.

  Finally, there is one more possibility.  Since the selective-Mott state must also undergo ordering instabilities at sufficiently low $T$, if only to relieve its finite $T=0$ entropy (when 
$V_{fc}(k)$ is irrelevant), the infra-red singular behavior can also be cut-off
by a direct instability to two-particle coherence.  This needs an extension of the above framework to explicitly include broken symmetry,
and we will present a specific scenario in a phenomenological way when we study phonon responses in the underdoped cuprates below.

\vspace{0.5cm}

{\bf Phonon Dynamics from INS Results}

\vspace{0.5cm}

  We now extend the above approach to study various features related to anomalous lattice dynamics in cuprates.  In particular, we 
will focus on $(i)$ inelastic neutron scattering data, which shows clear anomalies in phonon dispersions and lineshapes, $(ii)$ the
relative importance of the electronic bosonic and phononic contributions to the pairing glue, and $(iii)$ use of lattice dynamics
studies in underdoped cuprates as a test of hidden order in the pseudogap phase that emerges in underdoped cuprates.
 
  Inelastic neutron scattering (INS) studies provide valuable information on lattice dynamics: in particular, 
the phonon lineshapes, dispersions and linewidths can be extracted from INS data.  In the context of high-T$_{c}$ 
cuprates, such studies have been extensively performed over the years by various groups~\cite{reznik}.  Spurred on by the  need to derive quantitative estimates for relevance of electron-lattice coupling in HTSC, special attention
has been devoted to various issues.  These are

$(i)$ how does the character of lattice dynamics evolve from the strange metal near optimal doping, as one crosses 
over into the famed pseudogap (PG) regime in the underdoped (UD) regime?

$(ii)$ can changes in phonon spectra provide us with a handle on the important issue of discerning the microscopic nature
of the PG state?  Characterizing the PG state is an extremely controversial subject, with various competing scenarios 
extant in literature.  These range from preformed $d$-wave phase fluctuations without actual SC coherence~\cite{wang}, nematic instability~\cite{senthil}, circulating-current order~\cite{varma}, $d$-density-wave~\cite{sudip}, and stripes~\cite{tranquada}, among others.  
Appropriate smoking gun scenarios can greatly aid in distinguishing between these.  In any case, any plausible scenario
should be constrained by an all-important known fact: namely, that the PG regime with its 
possible novel symmetry-breaking (or lack thereof) must arise as an instability of the strange metal without Landau quasiparticles.  

$(iii)$ study of lattice dynamics in HTSC has long been confronted with a set of well-known, but ill-settled issues.  These
are:

$(1)$ in UD cuprates, halfway the Brillioun zone, the ’half-breathing’ Cu-O vibration
mode seems to suddenly dip down to a much lower frequency~\cite{reznik}.  The line-width reaches its maximum at
a wave-vector somewhat shifted from the frequency-dip
position, while it narrows at higher temperatures~\cite{fujita}.  A pronounced and little-understood asymmetry in the spectra for UD cases is also visible.  In addition, the anomaly has a narrow intrinsic peak width
as function of momentum transversal to the mode propagation direction.  These rather dramatic observations have been 
hitherto addressed within fluctuating stripe~\cite{zaneen} and $t-J$ model approaches~\cite{khaliullin}, but revisiting 
this problem is called for in view of new proposals regarding the order in the pseudo-gap phase.  This is because 
changes in phonon dynamics upon entering the PG phase must, in the final analysis, be tied down to possible electronic 
order (or lack thereof) in that phase.  

$(2)$ For cuprates, a rather weak EPI was found, which alone would not be sufficient to explain the
superconductivity. However, the calculated width of the half-breathing phonon
is an order of magnitude smaller than the reported experimental value, raising some
questions about the accuracy of the LDA in this context~\cite{reznik}.  In particular, this points to the need to 
include strong lifetime effects in the phonon propagator: the latter can only arise from incorporation of dynamical fluctuation effects in the electron-phonon interaction, and the latter is expected to be particularly unconventional in the strange metal (see below).

$(3)$ while signatures of polarons are purportedly seen in undoped cuprates, it is much more challenging to interpret 
manifestations of electron-phonon coupling in the strange metal.  This is because the very untenability of a low-energy
Landau quasiparticle description for the strange metal inevitably leads to non-trivial features in extracting such 
information when the one- and two-fermion propagators exhibit an infra-red branch-cut singular structure.  Specifically, 
observation of anomalous continuum response in the magnetic fluctuation spectrum in the strange metal implies that
associated features should manifest in phonon spectra.

  Thus, if we wish to extend the analysis of spin and charge fluctuations above to study phonon spectra, the interesting and relevant issue we are faced with is whether (and to what extent) the above issues can be understood within a theoretical scenario where phonons 
are coupled to a specific critical electronic continuum as worked out above.  More importantly, we also ask: {\it can we use 
changes in phonon dynamics upon entering the pseudo-gap phase as a tool that can offer an additional valuable insight
into novel order in the PG state?}  In this section, we address this issue, as
well as the related one of relative importance of intrinsic multi-electronic and phononic glues for SC, in some
detail.

  The free phonon propagator for a $D=2$ square lattice is

\be
D(q,\omega)=\frac{2\omega_{q}}{\omega^{2}-\omega_{q}^{2}}
\ee

We choose the electron-phonon coupling to reflect the fact that the coupling reflects
electron density modulation by dispersive half-breathing bond-phonons~\cite{khaliullin}: the coupling Hamiltonian
is

\be
H_{e-p}=\frac{1}{\sqrt{N}}\sum_{k,q}g(q)c_{k+q,\sigma}^{\dag}c_{k,\sigma}(b_{q}+b_{-q}^{\dag})
\ee
where
$g(q)=g\sqrt{sin^{2}(q_{x}a/2) + sin^{2}(q_{y}a/2)}$ and $q = (q_{x},q_{y})$. Formally, the renormalization
of the phonon propagator is caused by coupling to a particle-hole susceptibility, $\chi_{cc}(q,\omega)$ in the charge sector.
In a Fermi liquid metal, the latter is evaluated from the fully dressed electronic GFs in a RPA
summation involving the fully renormalized GFs and the bare phonon vertex. Effects beyond
HF-RPA, which result in a dressing of $g(q)$, do not qualitatively modify that picture. 
In the quantum critical metal, however, the finding that the density fluctuation spectrum has no pole
structure, but rather a branch-cut analytic structure in the infra-red (with a fractional exponent $\alpha_{c}$ distinct from The exponent $\alpha_{s}$ in the spin fluctuation spectrum) leads us to expect qualitatively
radical deviations from this traditional picture.  Thus, given a $\chi_{bb}(q,\omega)$ having similar infra-red branch-cut continuum form as above, we can now readily proceed to discuss the phonon spectra in our case.  The dressed phonon GF is, as usual related to the density susceptibility as
follows:

\be
D(q,\omega)=\frac{2\omega_{q}}{\omega^{2}-\omega_{q}^{2}-2g^{2}(q)\omega_{q}\chi(q,\omega)}
\ee
with $\chi(q,\omega)=[\chi_{0}^{-1}(q,\omega)+J(q)]^{-1}$ and $\chi_{0}(q,\omega)=C(q)T^{-\alpha'}h(\omega/T)$.  Here,
$h(\omega/T)=\chi_{0}(iC/2\pi T)^{\alpha'}\frac{\Gamma(\alpha')\Gamma(\frac{1-\alpha'}{2\pi}-\frac{i\omega}{2\pi T})}{\Gamma(\frac{1+\alpha'}{2\pi}-\frac{i\omega}{2\pi T})}$.  Here, $J(q)=2J(cosq_{x}a +cosq_{y}a)$
and $\Gamma(x)$ is the digamma function of $x$.  In the strange metal, $\alpha'=(1/\pi)$tan$^{-1}(U_{fc\uparrow\uparrow}/W) < 1$ is the anomalous exponent coming from the infra-red
singular electronic susceptibility as above, and $\chi_{0}$ and $C$ are adjustable parameters.  The phonon lineshape is 
now read off as the spectral function of the renormalized phonon propagator, i.e, it is $A_{ph}(q,\omega)=-\frac{1}{\pi}$Im$D(q,\omega)$, while the renormalized phonon dispersion is simply given by the equation $\omega^{2}-\omega_{q}^{2}-2g^{2}(q)\omega_{q}\chi(q,\omega)=0$.  Finally, the phonon damping, measured as the full-width at half-maximum of the phonon
lineshape, is simply the imaginary part of the phonon self-energy above.
  We present our results in Fig.~\ref{fig5},~\ref{fig6},~\ref{fig7}.  Several novel features, germane to extant data, stand out without the need to make any further assumptions about the electronic state:

\begin{figure}[h!]
\centering
\includegraphics[angle=0,width=0.96\columnwidth]{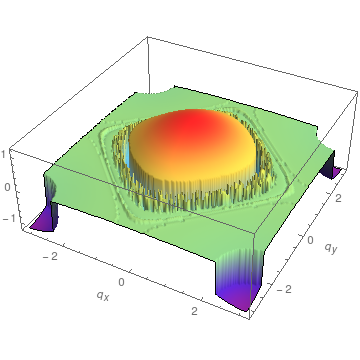}
\caption{Renormalized Phonon dispersion relations in the strange metal phase in the Brillouin zone, 
$\omega(q_{x},q_{y})$ at $\beta$ = 1000. Standard color function has been used for dispersion intensity (violet having least value and
red having highest value).}
\label{fig5}
\end{figure}

\begin{figure}[h!]
\centering
\includegraphics[angle=0,width=1.00\columnwidth]{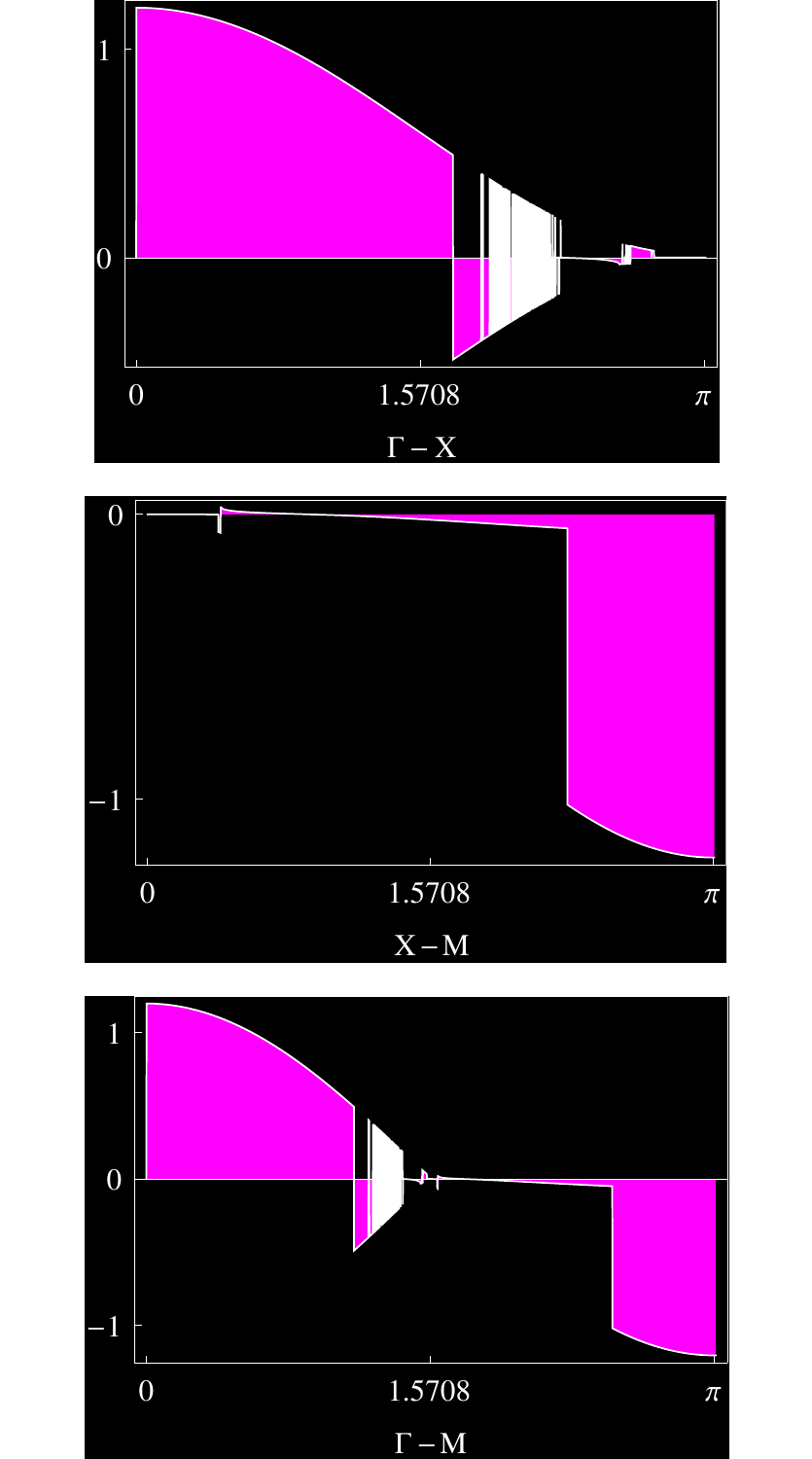}
\caption{Renormalized bond-stretching phonon dispersions along different high symmetry directions of the reduced Brillouin zone 
at $\beta$ = 1000.  The severe downward bending of the phonon dispersion midway the Brillouin zone along both ${\bf q}_{1}=(h,0,0)$ and ${\bf q}_{2}=(0,k,0)$ is in very good qualitative accord with data of Reznik {\it et al.}~\cite{reznik}, as explained in the text.}
\label{fig6}
\end{figure}
$(1)$ {\bf Phonon dispersions}  As mentioned above, in at least two families (LSCO and YBCO, the latter at doping levels 
unrelated to any putative stripe-like instabilities), the half-breathing dispersive bond-phonon mode shows anomalous softening half-way across the Brillouin zone, followed by an anomalous growth of the linewidth at a slightly different 
wave-vector.  This is particularly obvious in Figures (2),(4) of Reznik {\it et al.}~\cite{reznik}, where the phonon dispersion shows
a completely unanticipated and severe bending from its LDA counterpart.  While this should be interpreted as being caused
by coupling of the phonons with appropriate symmetry to electronic fluctuations, large damping and associated asymmetric 
lineshapes show that this likely involves strong coupling to electronic excitations that are themselves strongly damped.  
Our contention is that these electronic modes are precisely those associated with the anomalously damped electronic continuum derived above.  Our results bear out this expectation quite satisfactorily.  It is indeed quite satisfying 
that the phonon dispersion shows a sharp drop (actually, a jump in $\omega(q)$) precisely somewhat midway across the 
Brillouin zone, both along ${\bf q}_{1}=(h,0,0)$ and along ${\bf q}_{2}=(0,k,0)$, corresponding to Figs.(2),(4) of 
Reznik {\it et al.}  This feature arises simply due to coupling of the half-breathing phonon to an infra-red critical multielectronic continuum in our picture.  The attractive feature of the present work is that, apart from coupling to 
a collective fluctuation spectrum of (anomalously overdamped) fermionic origin, no other assumptions are needed.
To the extent that this anomalous fermionic fluctuation spectrum is also consistent with a host of unusual observations
in the strange metal, this accord lends additional support to the notion of an anomalous locally critical fluctuation spectrum in the strange metal.
\begin{figure}[h!]
\centering
\includegraphics[angle=0,width=0.96\columnwidth]{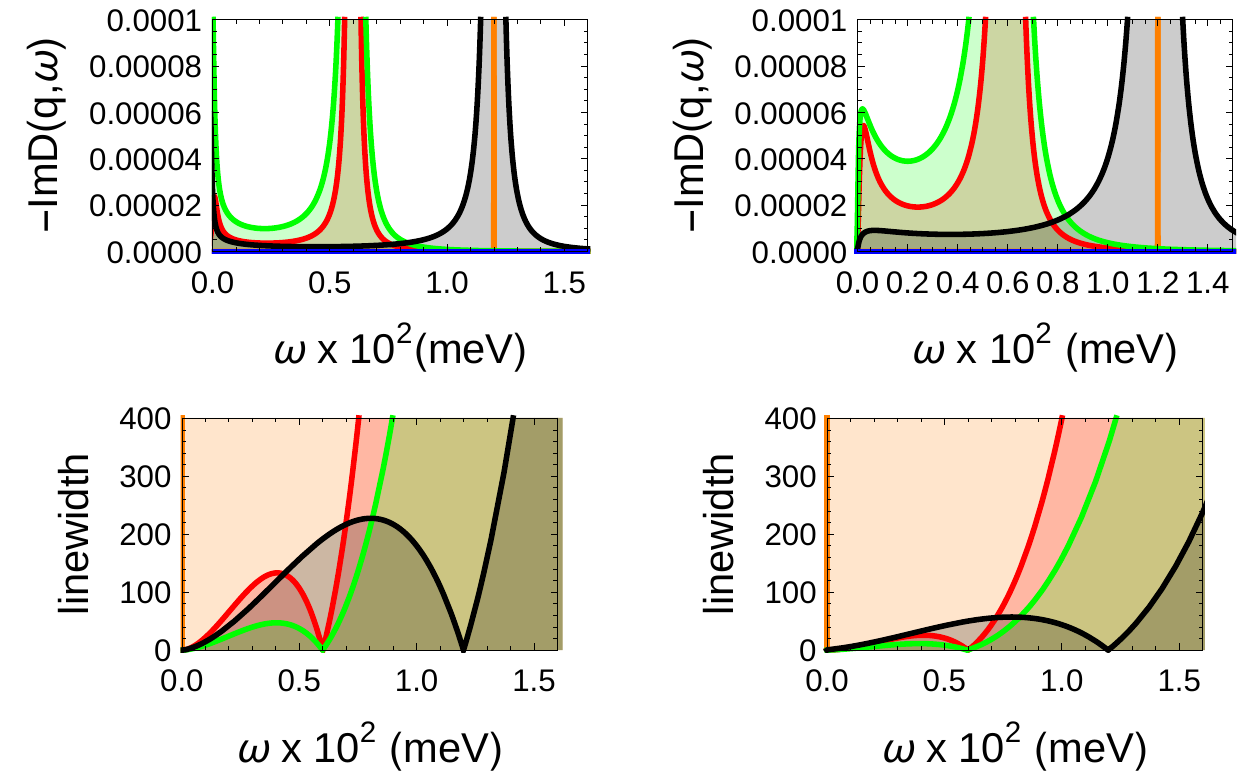}
\caption{Renormalized Phonon lineshapes and linewidths as functions of various ${\bf q}$ in the Brillouin zone.  \textcolor{orange}{(0,0)}\textcolor{red}{($\pi/2,0$)}\textcolor{purple}{($\pi,0$)}\textcolor{green}{($\pi,\pi/2$)}\textcolor{black}{($\pi,\pi$)}\textcolor{blue}{($\pi/2,\pi/2$)} of the reduced Brillouin zone at $\beta$ = 100 (left panel) and 1000 (right panel).  Asymmetry in the lineshapes as well as a ``Fano''-like peak 
at low energy is also clearly visible.}
\label{fig7}
\end{figure}
\vspace{0.5em}

$(2)$ {\bf Phonon Lineshapes}  Associated anomalous features are also visible in phonon lineshapes, also shown in 
Fig.~\ref{fig7}.  A noticeable asymmetry, centered around the (main) phonon peak at $60$~meV is visible.
In a Fermi liquid, such asymmetry is not expected.  More dramatically, an anomalous and incoherent shake-up feature is 
also found in our calculation around $\omega=0$, and the phonon spectral function is thus composed of $(i)$ a strong feature around the phonon frequency, shifted somewhat from its bare value and anomalously broadened, especially at lower $T$.  This is the direct effect of the phonon self-energy.  A direct prediction following herefrom is that the phonon linewidth should show appreciable damping at a wave-vector somewhat different from that where the dispersion shows the 
anomaly in $(1)$ above.  Thus, the direct effect of coupling to a critical multielectronic continuum is that the 
phonon lineshapes will also exhibit anomalies.  It would be very interesting to analyze the prediction of a specific
Fano-like shake-up feature in phonon lineshapes, but we remain unaware of systematic studies along these lines.  The key 
stumbling block seems to be the limit of resolution in actual INS studies: since the shake-up feature is very small
compared to the main phonon peak, it will probably be completely obscured by the background signal in reality.  But another key prediction, that of measurable asymmetry in the main phonon lineshape, could be more readily resolved with 
state-of-the-art facilities.  Appreciable phonon damping is also clearly seen in our results, and
 is qualitatively consistent with the large order-of-magnitude discrepancy 
between LDA estimates and neutron data for phonon linewidths~\cite{reznik}.

\begin{figure}[h!]
\includegraphics[angle=0,width=0.98\columnwidth]{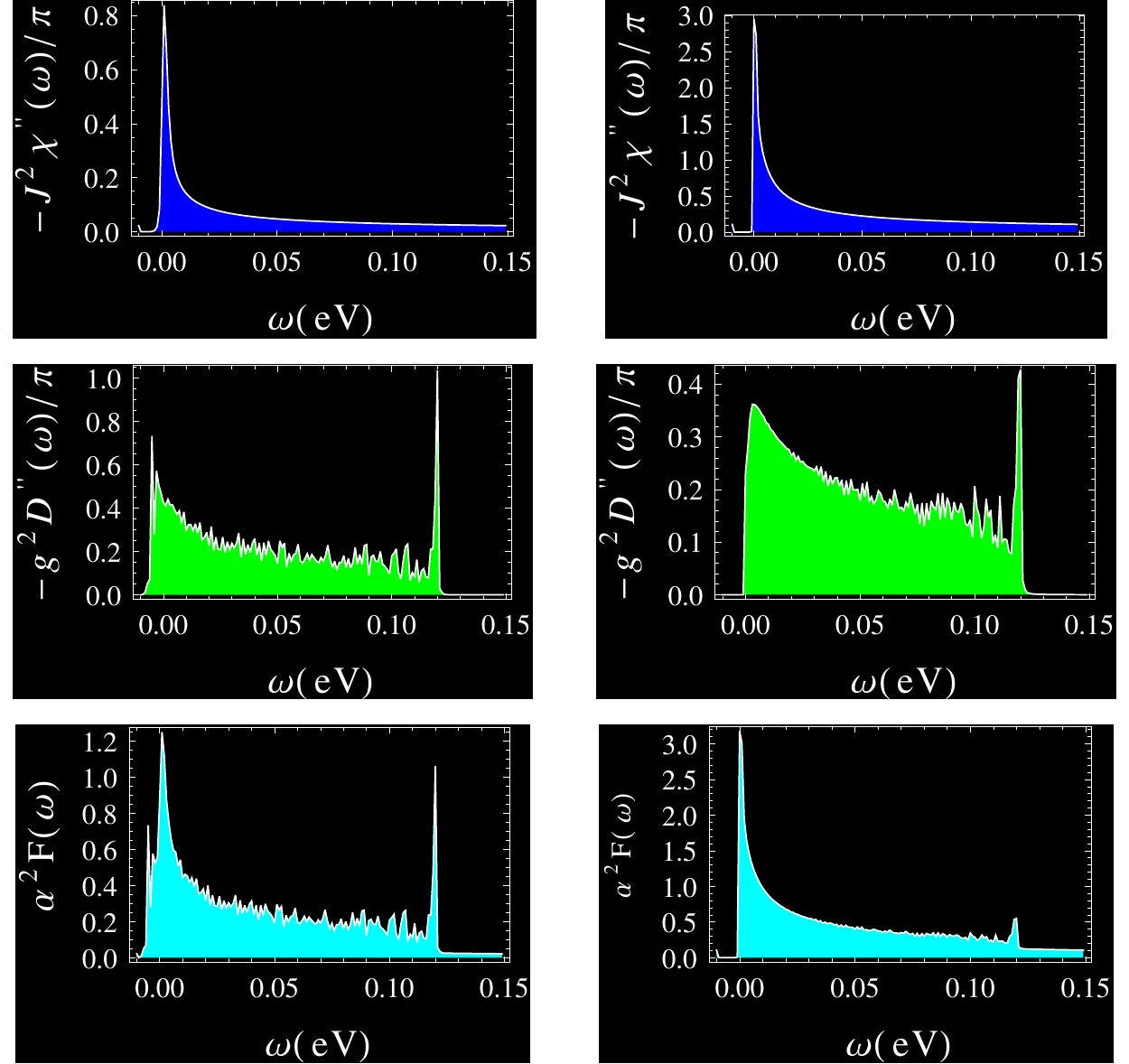}
\centering
\caption{Multi-electronic and bosonic contributions to the pair-glue functions, along with the total glue function at $\beta$ = 100 (left panel) and 1000 (right panel).  At low $T$, the contribution of the multi-electronic glue is an order-of-magnitude larger than that due to bond-stretching phonons, in good accord with quantitative estimations from pump-probe data (see text).}
\label{fig8}
\end{figure}

$(3)$ Finally, our results for the electronic susceptibility and phonon lineshape can be directly used to estimate the pairing glue function contribution from both processes.  The relative contributions from purely electronic and phononic
channels to the high-T$_{c}$ values in cuprates has been a subject of various detailed analyses~\cite{conte,reznik}.
  However, the relative importance of these processes has never, to our best knowledge, been considered within the 
strange metal hypothesis, though cluster-DMFT studies indeed go a long way toward showing the dominant importance of a purely electronic pairing glue.  Clearly, this is an important constraint on theoretical scenarios, and here, we present
our attempt to fill in this breach.

  Within our analysis above, the pair glue function is simply a sum of two contributions, $(i)$ from the purely electronic collective modes, written as $\alpha_{e}^{2}F_{e}(\omega)=J^{2}\sum_{q}$Im$\chi(q,\omega)$, and a phononic part, which is 
simply $\alpha_{p}^{2}F_{p}(\omega)=\sum_{q}g^{2}(q)$Im$D(q,\omega)$.  
In Fig.~\ref{fig8}, we show these contributions and the total pair glue function, which exhibits several interesting features.  

  First, $\alpha_{e}^{2}F_{e}(\omega)$ at low $T$
clearly exhibits the singular branch-cut continuum form $\simeq \omega^{-\mu}$ inherited from the strange metal.  This 
is precisely the contribution expected from a quantum critical fluctuation spectrum, whose origin (in our picture) is 
the critical (incoherent) continuum that arises due to the underlying inverse orthogonality catastrophe in the
selective-Mott state in the fermionic (DMFT) theory.  Interestingly, though the phononic contribution is not negligible, it
 is clearly much smaller (almost by an order of magnitude) than the purely electronic one.  Though a full theoretical 
estimate of these glues to $T_{c}$ must involve treatments beyond the Eliashberg scenario~\cite{moon}, we can 
readily make qualitatively reasonable order-of-magnitude estimates using the above results.  However, one needs to take into account the fact that having 
$\lambda_{e}(\omega)=\alpha_{e}^{2}F_{e}(\omega)/\omega\simeq J^{2}$Im$\chi(\omega)/\omega$ makes this case qualitatively different from a BCS-Eliashberg one, where $\lambda(\omega)$ is constant at low energy.  Specifically, retardation effects are much more subtle when the pair glue is dominated by a massless (critical) electronic boson.  Fortunately, the singular-glue case has been recently analyzed by Moon {\it et al.}.  The pairing scale, denoted $T_{p}^{el}$, increases monotonically as the quantum critical regime is approached from the disordered side.  However, the contribution to the
actual SC $T_{c}^{el}$ from the electronic continuum, estimated following Moon {\it et al.},
 is given by $T_{c}^{el}\simeq E_{F}$. $e^{-\pi/2\sqrt{\lambda}}$, enhanced 
above its BCS value.  Thus, this strong coupling view gives two different scales: one associated 
with pair formation without pair coherence (at $T_{p}$) and the second with 
long-range pair coherence ($T_{c}^{el}<T_{p}$), where actual SC occurs.  
However, the contributions from phonons can be qualitatively estimated within 
a more traditional BCS-Eliashberg formalism, since $F_{p}(\omega)$ does not 
exhibit low-energy singular behavior.  Inserting the order-of-magnitude values 
from Fig.~\ref{fig8}, it is easy to convince oneself that the contribution of the electronic (``bosonic'') 
continuum to $T_{c}$ is about an order-of-magnitude higher than that of phonons.  
Thus, our findings are also in full accord with other theoretical and 
experimental~\cite{conte} estimations on relative importance of electronic and phononic glues in cuprates.

\vspace{1.4cm}

{\bf Pseudogap Regime}

\vspace{0.4cm}
  Our treatment has hitherto relied on the existence of strange metallicity, and thus restricted to the near-optimally doped regime.  Upon underdoping, even more drastic deviations from Fermi liquid theory in thermal and transport responses
characterize the famed pseudogap regime.  

$(1)$ inelastic neutron scattering clearly reveal sizable $1D$-like anisotropy in magnetic excitations around the AF ordering wave-vector ${\bf Q}=((\pi/2,\pi/2)$, in a magnetically disordered phase below $150$~K in the UD phase.  That this anisotropy is most likely associated with an intrinsic electronic instability is suggested by the observation that it greatly exceeds estimates based on purely structural anisotropy, and, more crucially, by the finding that it follows the $T$ and $x$ dependence of the inplane resistivity anisotropy~\cite{keimer}.

$(2)$  more recent structural measurements by Fujita {\it et al.}~\cite{fujita} also finds a $d$-wave modulation of the 
charge density distribution in the UD cuprates.  
Interestingly, both, this modulation and nematic signatures in $(1)$ above, 
appear to vanish precisely around the same doping close to optimal doping, 
making them attractive candidates for the hidden order in the UD cuprates.  
It must be emphasized that both these orders can co-exist on purely symmetry 
grounds, and, indeed, that one follows the other. 

$(3)$ simultaneously, however, Nernst effect data~\cite{ong} appear to indicate a strongly fluctuating, preformed cooper pair phase without actual superconductive order.  This has been studied in detail by various groups~\cite{kivelson,scalapinolee}
with good success.  

  $(1)-(3)$ pose an additional set of questions.  How can we conceive of $(1),(2)$ arising as direct instabilities of the
strange metal found around optimal doping?  Are they in irreconcilable conflict with $(3)$, which has also been successfully used to rationalize a variety of data?

 Within our approach as described above, we proceed as follows. 
Since the local strange metal found within DMFT has a finite residual entropy, this state must
eventually find a way to relieve this entropy as $T\rightarrow 0$.  If one-electron hybridization were eventually relevant, quenching the entropy would necessarily involve eventual screening of the selectively Mott-localized magnetic moment via the local Kondo effect (this is indeed what happens in the EPAM for small enough $U_{fc}$).  However, 
since $V_{fc}(k)$ is irrelevant in the strange metal, the system must quench its residual entropy by generating 
a long-range ordered state via a two-particle instability, directly from the strange metal.  In analogy with what happens in coupled $D=1$ Luttinger
liquids and following earlier ideas of Anderson~\cite{pwa}, we have proposed~\cite{laad} that increasing relevance
of inter-site correlations between neighboring impurities of the DMFT problem generates an effective residual 
and non-local two-particle interaction in the local version of the strange metal.  As in the coupled $D=1$ chains case, this allows for a direct instability of 
the local critical metal to ordered states.  As worked out before, this residual two-particle interaction is

\be
H_{res} \simeq -\frac{V_{fc}^{2}}{U_{fc}}\sum_{k,k'}\gamma(k)\gamma(k')(b_{k,\sigma}^{\dag})a_{k\sigma}a_{k'\sigma'}^{\dag}b_{k'\sigma'} + h.c)
\ee
which is also 

\be
H_{res} \simeq -\frac{V_{fc}^{2}}{U_{fc}}\sum_{k,k'}\gamma(k)\gamma(k')[b_{k\sigma}^{\dag}(\delta_{kk'}\delta_{\sigma\sigma'}-a_{k'\sigma'}^{\dag}a_{k\sigma})b_{k'\sigma'} + h.c]
\ee
and has a separable form in $k$-space.
Within the local approximation, it is now fully legitimate to decouple $H_{res}$ (which scales like $1/D$) in a Hartree-Fock approximation in both particle-hole (p-h) and particle-particle (p-p) channels.  The result is

\be
H_{res}^{mf}= -J'\sum_{k}[\Delta_{ph}(k)b_{k\sigma}^{\dag}a_{k\sigma} +\Delta_{pp}(k)b_{k\sigma}^{\dag}a_{-k\sigma'}^{\dag}+ (1-\langle n_{k,a,\sigma}\rangle)n_{k,b,\sigma}]
\ee
and, since both $\Delta_{ph}(k),\Delta_{pp}(k)$ arise from the same set of $a,b$ fermionic states involved in $H_{res}$,
it follows that they necessarily represent competing p-h (density-wave) and p-p (superconductive) orders.  Additionally, the last term in $H_{res}$ above results in electronic anisotropy mfrom the outset, 
yielding an electronic nematic that co-exists with $\Delta_{ph}\neq 0$.  It is crucial to notice that, with $\gamma(k)=$(cos$k_{x}-$cos$k_{y})$, both p-h and p-p instabilities are of $d$-wave type.  The p-h order parameter is the average $\Delta_{ph}(k)=\langle \gamma(k)a_{k\sigma}^{\dag}b_{k\sigma}\rangle$ while the p-p order
parameter is $\Delta_{pp}(k)=\langle\gamma(k)b_{k\sigma}^{\dag}a_{-k,-\sigma}^{\dag}\rangle$.  
The first represents a $d$-wave charge density modulated order.  
As advertized above and previously noted by Kee et al., such
a $d$-density modulation, 
expressed as $\Delta_{d,ch}(q)=\sum_{k}\gamma(k)b_{k+q,\sigma}^{\dag}a_{k\sigma}$, will also necessarily follow a finite
$\Delta_{ph}(k)$ (and vice-versa) on purely symmetry grounds.  Thus, starting from the strange metal, we have derived an effective residual
interaction which exposes both, the instability to a $d$-wave superconductor, as well as to a competing
p-h or excitonic instability to a state with a $d$-wave density modulation and intra-cell nematicity, in qualitative
accord with recent structural analyses~\cite{fujita}.

\vspace{0.4cm}
\begin{figure}[h!]
\centering
\includegraphics[angle=0,width=0.98\columnwidth]{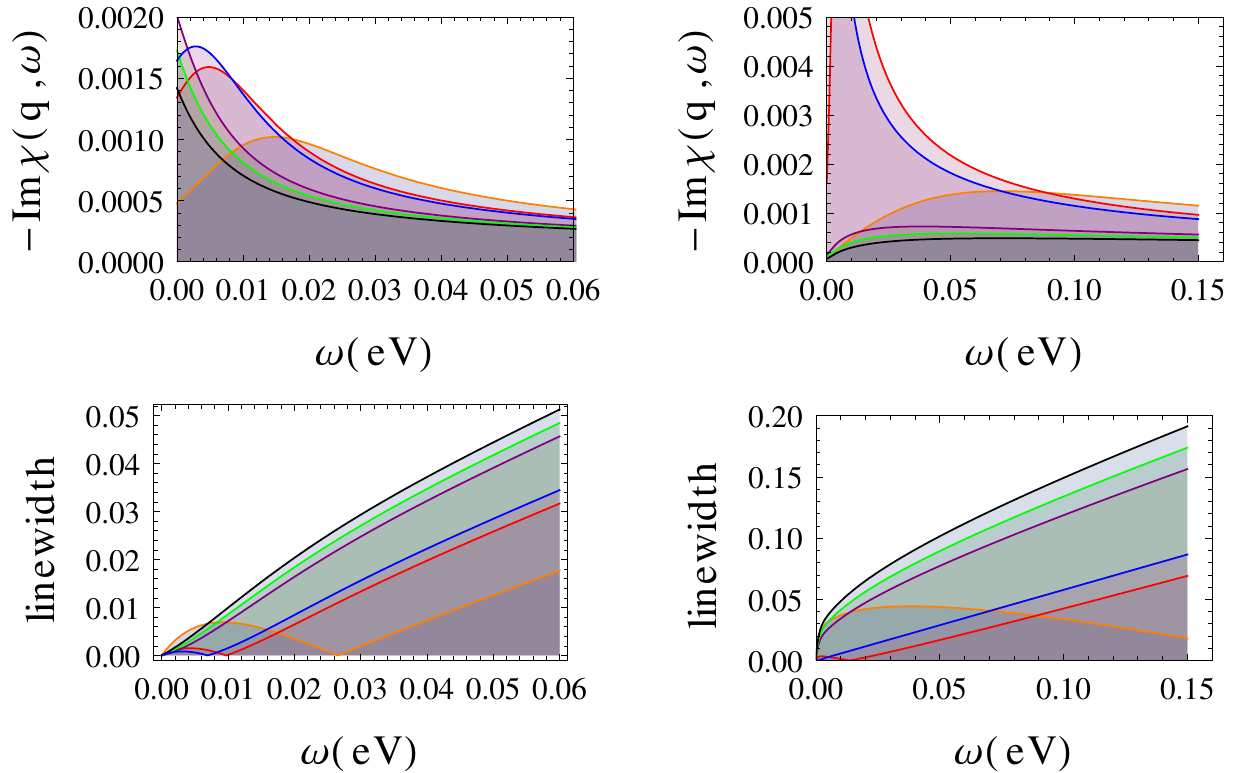}
\caption{The dynamical spin susceptibility and spin-fluctuation linewidths for the underdoped cuprates at various ${\bf q}$ in the Brillouin zone \textcolor{orange}{(0,0)}\textcolor{red}{($\pi/2,0$)}\textcolor{purple}{($\pi,0$)}\textcolor{green}{($\pi,\pi/2$)}\textcolor{black}{($\pi,\pi$)}\textcolor{blue}{($\pi/2,\pi/2$)} of the reduced Brillouin zone at $\beta$ = 100 (left panel) and 1000 (right panel).  Phenomenological introduction of a nematic-plus-$d$-wave modulation electronic order suppresses the infra-red singularity in spin-fluctuations in a momentum-dependent way.}
\label{fig9}
\end{figure}

\begin{figure}[h!]
\centering
\includegraphics[angle=0,width=0.96\columnwidth]{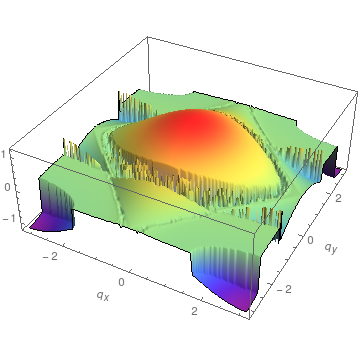}
\caption{Renormalized Phonon dispersion relations in the pseudogap phase in the Brillouin zone, 
$\omega(q_{x},q_{y})$ at $\beta$ = 1000.  
Clear momentum-dependeent anisotropy, along with clear anomaly in the 
dispersion along $q_{y}$ in good accord with data of Pintschovius {\it et al.}~\cite{pint} is clearly visible.
Standard color function has been used for dispersion intensity (violet having least value and
red having highest value). }
\label{fig10}
\end{figure}

  With this insight, we now return to the question posed before: can we use the changes in phonon spectra in the PG phase
to gain insight into possible novel electronic order (or lack thereof) in UD cuprates?  
From the above analysis, it is clear that, if a hidden order related to $\Delta_{ph}(k) \neq 0$ indeed develops in the PG state, now to be interpreted as a direct instability of the strange metal, this will immediately lead to modification of the inter-site
and multi-band two-particle interaction in an anisotropic way, thanks to the $d$-wave form factor in $H_{res}$.  At the
most basic and non-self-consistent level, this directly implies an effective $d$-wave contribution to the two-particle sector (self-consistency in a fully microscopic fermionic model will, of course, affect the result quantitatively but not
qualitatively).  Adopting a purely phenomenological view, we introduce such a term by hand in our approach: having $d$-wave density modulation plus electronic nematic order will directly introduce anisotropy in the dispersion as $J_{eff}(q)=J($cos$q_{x}+$cos$q_{y})+\delta J($cos$q_{x}-$cos$q_{y})$, with $\delta J$ being linearly related to $J'$ above.  Recalculating the neutron scattering lineshape and phonon dispersion as well as lineshapes above allows us to analyze the changes in these responses and gauge
the extent to which they offer a rationalization of features mentioned before.  We show the re-computed results, representative for the PG phase, in Fig.~\ref{fig9},~\ref{fig10}.

{\bf Inelastic Neutron Scattering Lineshape}
  Our main results for $\chi''(q,\omega)$ are shown in Fig.~\ref{fig9}.  The first observation is that the infra-red singular feature seen for the strange metal is immediately cut-off by the appearance of the $d$-wave pseudogap scale (which is of order $j\delta$ on the scale of $50$~meV in actual UD cuprates), the latter
now being introduced by a finite $\Delta_{ph}(k), \Delta_{d,ch}$ as above.  As also expected, the $d$-wave nature inherent in $H_{res}$ leads to significant variation with ${\bf q}$, and the resulting anisotropy in the magnetic fluctuation spectrum with respect to $q_{x}$ and $q_{y}$ is 
clearly reflected in the results.  Specifically, the response around ${\bf Q}=(\pi,\pi)$ remains sharply peaked, but significant suppression of magnetic spectral weight for ${\bf q}\neq {\bf Q}$ stands out as a manifestation of anisotropic
spin-gap formation.  Concomitantly, increased damping, resulting in even less well-defined spin fluctuation modes, 
except for ${\bf q}={\bf Q}$, is also visible.  These are direct consequences of the proposed $d$-wave nematic-plus bond-order, 
which spontaneously breaks the discrete four-fold $C_{4v}$ rotation symmetry.  
It is important to point out, however, that all we need to obtain such 
a result is a finite intra-cell nematicity, but not necessarily a true long-range nematic 
order (sometimes dubbed fluctuating order in literature).  
So strictly speaking, this accord in our phenomenology cannot distinguish 
between strongly dynamically fluctuating nematic correlations and nematic order, 
but it is not inconsistent with true electronic nematic order either.  More
analysis is thus needed to clinch this issue.

\begin{figure}[h!]
\centering
\includegraphics[angle=0,width=1.0\columnwidth]{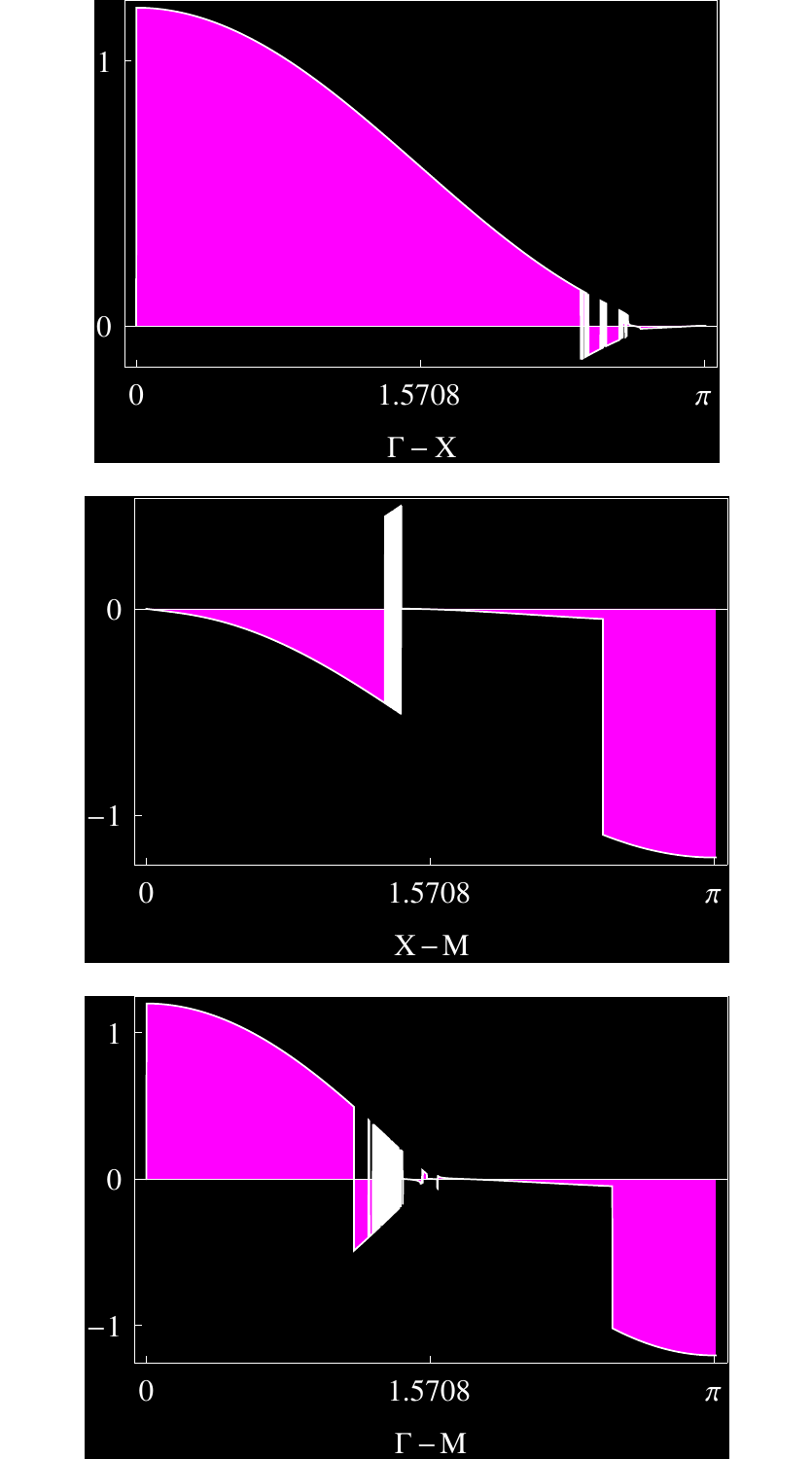}
\caption{Momentum selective phonon dispersions along different high symmetry directions of the reduced Brillouin zone 
at $\beta$ = 1000.  
Once more, the phonon dispersion shows severe downward bending at 
a wave-vector somewhat less than half-way the Brillouin zone as one moves along the path $\Gamma-M-X$ in the Brillouin zone.}
\label{fig11}
\end{figure}

\begin{figure}[h!]
\centering
\includegraphics[angle=0,width=0.98\columnwidth]{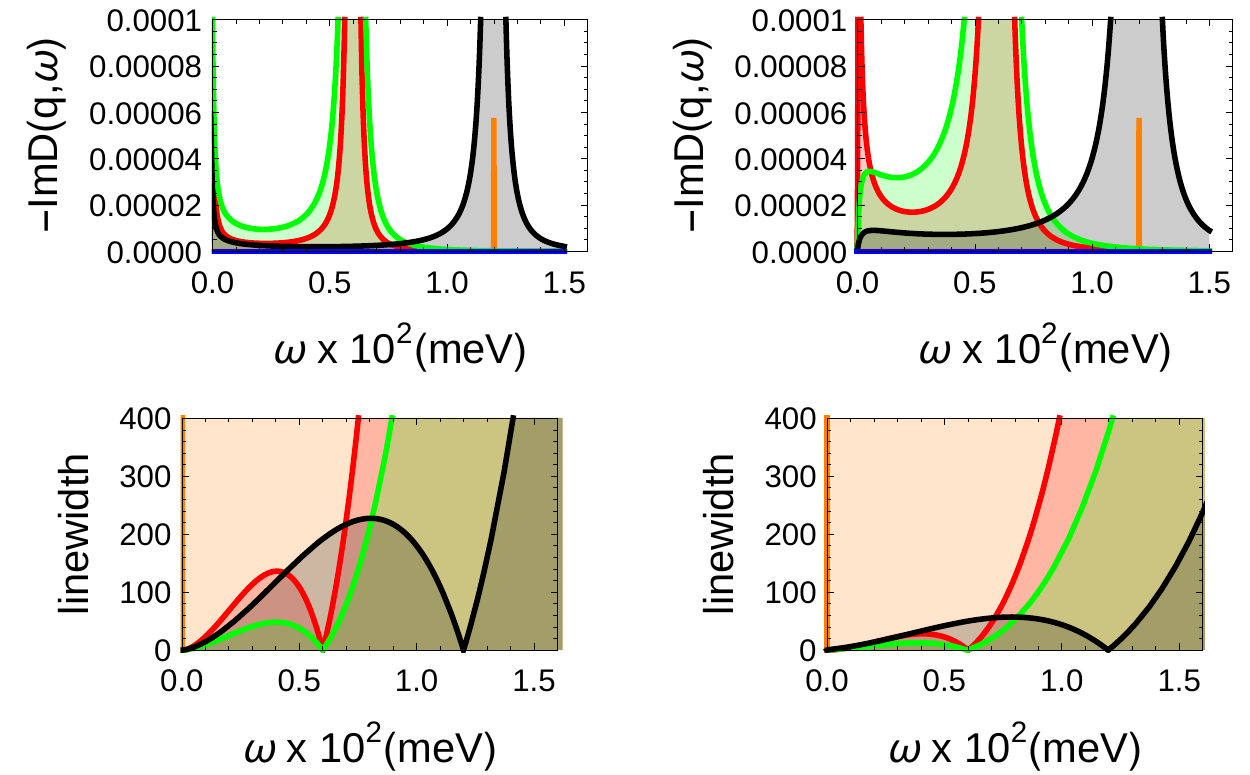}
\caption{Renormalized Phonon lineshapes and linewidths at different ${\bf q}$ points \textcolor{orange}{(0,0)}\textcolor{red}{($\pi/2,0$)}\textcolor{purple}{($\pi,0$)}\textcolor{green}{($\pi,\pi/2$)}\textcolor{black}{($\pi,\pi$)}\textcolor{blue}{($\pi/2,\pi/2$)} of the reduced Brillouin zone at $\beta$ = 100 (left panel) and 1000 (right panel).}
\label{fig12}
\end{figure}

\vspace{0.5cm}
{\bf Phonon Spectra in the $d$-PG state}
Concomitantly, phonon dynamics is also drastically modified by emergence of $d$-wave p-h instabilities in the PG phase.
We show the phonon dispersions Fig.~\ref{fig10},~\ref{fig11}, lineshapes, and linewidths Fig.~\ref{fig12} 
assuming finite $\Delta_{ph}(k)$.  In particular, we focus on the change in phonon 
dispersion within the scenario of an instability to $d$-wave 
nematic-plus $d$-wave density modulation.  The Fig.~\ref{fig10} clearly shows that the 
main effect of this instability is to introduce a clear anisotropy
in phonon dispersions.  Since $\delta j>0$, the dispersion of magnetic excitations (neglecting lifetime effects coming from the anomalous spin self-energy) will be $J(q_{x},q_{y})=(J+j\delta)$cos$q_{x} +(J-j\delta)$cos$q_{y}$.  This means
magnetic excitations are more dispersive along the $a$-axis and less so along $b$: this is a consequence of the $d$-wave
form of the residual interaction derived above.  One thus expects that the bond-stretching phonon mode will couple more
efficiently to electronic excitations along $b$, and this is fully borne out by our results.  Remarkably, we find a 
clear anomaly in the phonon dispersion along $b$ but not along $a$.  It is very interesting to observe that the anomaly
occurs in a direction perpendicular to the one along which electronic excitations are more dispersive, a finding fully 
consistent with that of Pintschovius {\it et al.}~\cite{pint}.  
As they have emphasized, this is inconsistent with a stripe scenario.  
However, we have now shown that it is qualitatively fully consistent with having 
intracell nematic-plus $d$-bond charge density modulated order in the UD phase.  Thus, our results offer an attractive possibility to distinguish between different scenarios for the hidden order of the PG phase by appealing to analysis of phonon spectra.  In the specific context of cuprates, the accord we find is corroborating evidence in favor of a coupled $d$-nematic-plus $d$-bond charge density modulated order in the UD cuprates, and reconciles phonon spectra with recent structural findings~\cite{fujita}.

{\bf What about Preformed-Pairs}?

    How does one reconcile extensive observations of preformed cooper-pair regime with findings above?  Extensive evidence of pre-formed cooper paired bad-metallic state 
is well documented in the PG phase.  Let us now discuss a possible resolution of this seemingly irreconcilable conflict
within our idea.  Focussing on the form of $H_{res}$ or $H_{res}^{mf}$, we see a remarkable emergent symmetry that may hold the key to our argument.  Introducing a $s$-wave spin-singlet pairing operator, $\Delta_{s,sc}=\frac{1}{\sqrt{2}}\sum_{k,\sigma}\sigma b_{-k,-\sigma}b_{k\sigma}$, we easily see that it rotates $\Delta_{d}$ above into $\Delta_{d,sc}$:

\be
[\Delta_{s,sc},\Delta_{d}]=\Delta_{d,sc}
\ee

  The crux is now that short-range correlations associated with $\Delta_{s,sc}$ also appear in the PG phase, thanks to the two-particle couplings
 between $a,b$ fermions in $H_{res}$ (remember that the sum is over both $a,b$ orbital indices).  
So having finite intra-cell nematic correlations in the PG phase immediately
implies, from the above Eq., that this phase will also be characterized by finite $d$-wave pair correlations, but without
actual phase coherence.  A similar situation also obtains in the FL${*}$ scenario~\cite{senthil} in the context of a fractionalized phase where a spin liquid asymptotically decouples from metallic fermions.  By itself, this observation is thus not at all inconsistent with the observations of preformed
pair-fluctuation signatures in the PG regime.  It provides a natural reconciliation of observed preformed $d$-wave pairing with the proposal of a ``hidden'' intracell Ising-nematic-plus $d$-wave charge modulation order in the PG phase.

\vspace{0.5cm}

{\bf Electronic and Phononic Pair Glues}

\begin{figure}[h!]
\centering
\includegraphics[angle=0,width=0.98\columnwidth]{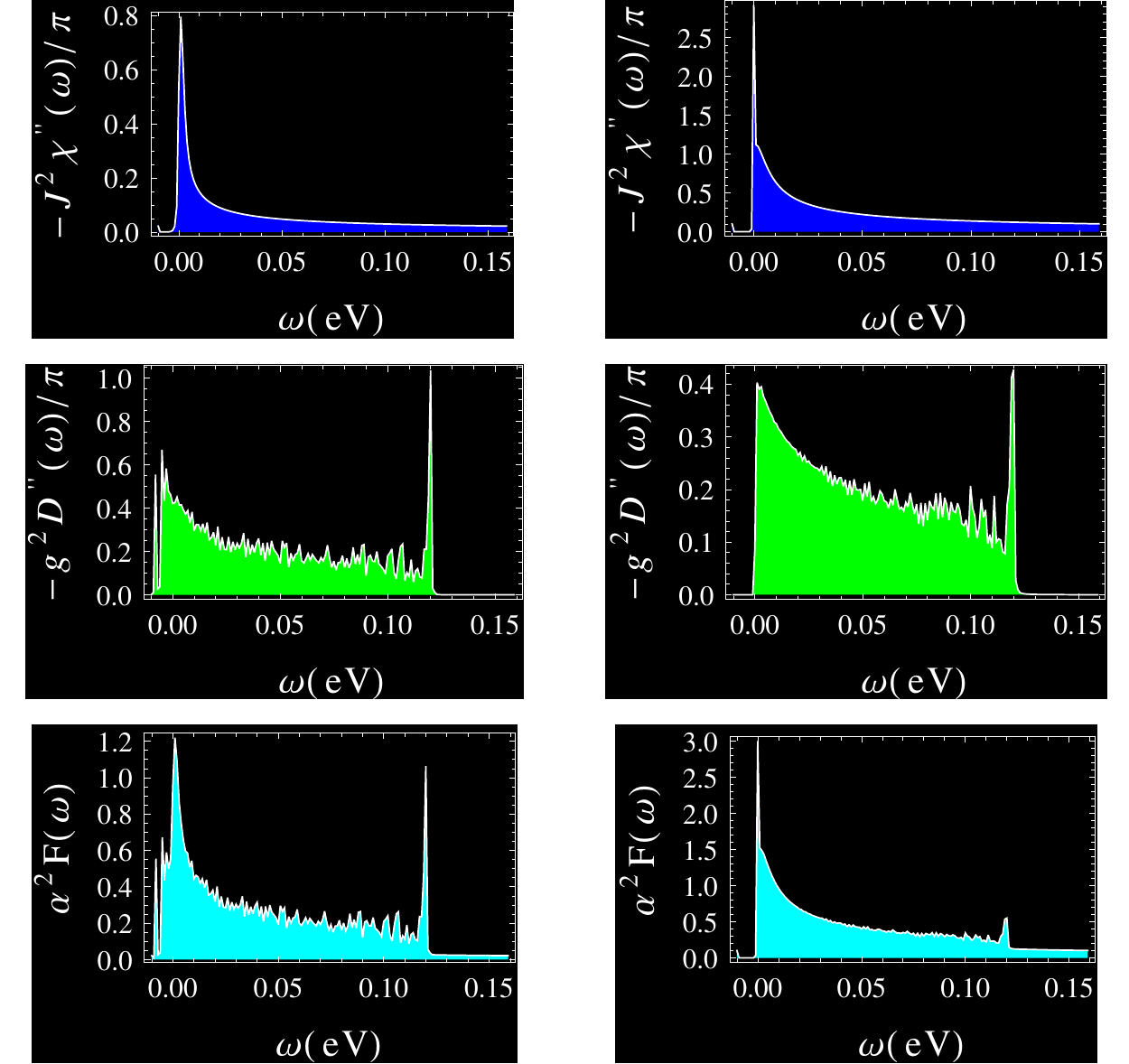}
\caption{Multi-electronic and phononic contributions, and the total glue function for the 
underdoped case at $\beta$ = 100 (left panel) and 1000 (right panel). 
Clear reduction in the critical multi-electronic occurs as a result of depletion of 
low-energy spectral weight, compared to the case for the strange metal. 
Reduction of SC $T_{c}$ upon underdoping could be driven by this effect. }
\label{fig13}
\end{figure}

\vspace{0.5cm}
  The changes in the magnetic fluctuation and phonon spectra upon emergence of 
intracell nematic-plus $d$wave density modulation order in UD cuprates must affect 
the total glue function, $\alpha^{2}F(\omega)=J^{2}$Im$\chi_{e}(\omega)+g^{2}$Im$D(\omega)$ defined before.  
That this is indeed the case is shown in Fig.~\ref{fig13}.  As one would expect, the cut-off of
the infra-red divergence in $\chi_{e}(q,\omega)$ in the PG phase cuts-off the corresponding feature in the glue function
contribution from electronic bosonic excitations.  By itself, this directly implies a reduction in the SC $T_{c}$ upon
under doping.  The underlying reason for this is clear from the observation that intracell 
nematic-plus $d$-charge density modulation competes with $d$-wave SC, since both instabilities 
arise from the same set of carriers, via the
residual interaction term.  It must be mentioned that though we have been able to rationalize the reduction of $T_{c}$ 
within our semi-phenomenological approach, a definitive resolution of this issue awaits a fully microscopic calculation
including $H_{res}$ in the corresponding fermionic model within DMFT or cluster-DMFT approaches.  This remains out of scope
 of the present study.  However, it is encouraging that a nematic instability as proposed here is indeed found in 
recent cluster-DMFT studies of the $D=2$ Hubbard model~\cite{cDMFT-nematic}.  
In view of our analysis, an attractive feature is that such a 
hidden order in the PG phase arises as a natural instability of the strange metal via the preferential dominance of intersite
 pair-hopping interactions in the p-h channel.

\vspace{0.5cm}

{\bf Discussion and Implications}  
\vspace{0.5cm}

  What distinguishes our view from extant ones?  In itinerant HMM views, the
spatio-temporal build-up of AF spin correlations near a $T=0$ AF QCP can, 
under certain conditions, give $\chi_{loc}(\omega)\simeq \omega^{-\eta}$ with
$\eta=1/2,1/3$~\cite{moon}.  Alternatively, in a Kondo-RKKY formalism, one
finds $\chi_{loc}(\omega) \simeq \omega^{-\gamma}$ with an interaction-dependent
 $\gamma$~\cite{si}.  In both cases, however, this behavior is intimately 
tied down to proximity to a $T=0$ AF QCP, even though it is accompanied by 
breakdown of the Kondo effect by competing intersite RKKY couplings in the 
latter case.  In stark contrast, similar behavior found in our selective-Mott
mechanism is not necessarily associated with proximity to any $T=0$ 
AF criticality:
the only quantum criticality involved here is that of proximity to the $T=0$
end-point of the line of first-order (liquid-gas universality) selective-Mott
transition: this implies a divergent compressibility and leads to destruction of 
Fermi liquid theory by coupling fermions to the associated soft collective 
electronic excitations.  What we
have shown here is that the inverse orthogonality catastrophe that acquires a rigorous 
meaning in the local problem of DMFT for the selective metal also provides, 
remarkably enough, a natural explanation for the anomalous scale-invariant 
magnetic fluctuations of the strange metal without having to invoke 
proximity to any $T=0$ AF QCP.  Further, since DMFT can yield a strange metal 
phase for a range of parameters in the EPAM, such anomalous spin correlations 
can exist over a range of tuning parameters in practice.  This is an 
attractive option when one considers cases like the optimally 
(hole)-doped cuprates and $\beta$-YbAl$_{4}$, which show features consistent 
with a quasi-locally critical $\chi(q,\omega)$ without any 
obvious proximity to $T=0$ AF order.  On the other hand, evidence for Fermi surface reconstruction 
(FSR) in both the above cases~\cite{gil,coleman} supports our argument 
linking these anomalies with selective-Mottness, where such FSR must 
indeed occur.
Though such FSR can also be rationalized within HMM and Kondo-RKKY views, they
must be associated with the $T=0$ AF QCP in those cases, as mentioned before.
In our case, the FSR involves a topological aspect of Mott
physics, namely, penetration of zeros of $G_{bb}(k,\omega)$ in EPAM to where
poles would have been in the normal metal~\cite{imada}, and has nothing to 
do with the FSR involving AF order.  However, in cases where the unquenched local moments in the OSMP
can (as mostly happens) order, the FSR will coincide with the magnetic instability.  Since it is
ubiquitous for spins to order in (insufficiently geometrically frustrated) real cases like cuprates 
and $f$-electron systems, it is hard to disentangle the roles of OSM physics vis-a-vis AF order
in the FSR that is widely known to occur near QCPs in those cases.  

  The anomalous spin- and charge-fluctuation continuum responses we find are an attractive and novel choice 
for an intrinsically multi-electronic glue that can induce an 
unconventional superconductive (USC) coherent state.  Indeed, suggestions for 
unconventional $d$-wave SC due to diverging AF spin correlations with an
effective dynamic pairing interaction of the form 
$V_{eff}\simeq J^{2}\chi(\omega)\simeq 1/\omega^{\eta}$ exist~\cite{moon}.  
Once again, in contrast to views where such a bosonic glue is 
associated with quantum critical fluctuations linked to destruction of AF 
order, the multiparticle 
incoherent glue function proposed here bears an intimate connection to breakdown of LFL theory at an OSMP.   It is intrinsically quantum critical in the sense of being a two-particle 
continuum that is inherently unstable to a $T=0$ USC state.  Finally, in contrast to a 
normal BCS instability, involvement of selective-Mott physics in emergence 
of local critical features (implying 
that both incoherent metallic and Mott localized fermions participate in 
pairing via 
$H_{res}$) naturally means wild pair-phase fluctuations from number-phase
uncertainty principle, suppression of the mean-field SC $T_{c}$ while 
preserving the pair amplitude, and thus strong preformed pair fluctuation 
regime above $T_{c}$: thus, this is, as expected, an intrinsically 
strong-coupling (non-BCS-Eliashberg type~\cite{moon}) instability of the 
strange metal.  Actual computation of the instability is much more involved in 
this case, owing to the fact that the pairing kernel requires the full form of 
$\chi({\bf q},\omega)$ over the entire energy range: this is out of scope of 
our analytic approach.

  There are more ways to further elucidate the unconventional nature of the 
underlying quantum criticality in our 
case.  In the selective-Mott state, below the selective-Mott gap, 
we find an emergent local Z$_{2}$
symmetry $[$exp$(i\pi n_{i,b}),H]=0$ for  all $i$, arising from $[n_{i,b},H]=0$
for all $i$.  This is an effective low-energy symmetry in the
$b$-fermion sector in the OSMP.  This feature is reminiscent of what happens in the celebrated Kitaev
model~\cite{kitaev}, where a Z$_{2}$ symmetry is exact rather than emergent, and maybe 
it is now no surprise that extreme short-range spatial correlations
 and X-ray-edge related behavior in $\chi(q,\omega)$ rigorously
characterize spin correlations there as well~\cite{baskaran,moess}.  However, in 
this case, infra-red power-law singular behavior results in stark contrast to 
the Kitaev case, where the Dirac fermion spectrum suppresses singular behavior.  Nevertheless, there are deeper analogies and connections with 
confinement-deconfinement (C-DC) transition(s): in our case, 
lack of $a$-fermion quasiparticles requires Mott localization of $b$-fermions,
while itinerant $b$-fermions imply a correlated FL metal.  In the 
language of fermions coupled to a fluctuating Z$_{2}$ gauge field as defined above, the former 
corresponds to a quasistatic gauge field, while the latter corresponds to the 
gauge field fluctuations acquiring dynamics with a timescale set by 
$\hbar/k_{B}T_{coh}$ (recollect that $T_{coh}$ is finite only when $V_{fc}(k)$ is relevant at one-particle level).  The exotic confinement-deconfinement (quantum) criticality that should be 
linked to this is thus 
once again - unsurprisingly - related to relevance or otherwise of $V_{fc}(k)$: As long as $V_{fc}$ remains irrelevant
(this can well be the case upto very low $T$ in practice, and often pre-empted 
by direct instabilities of the incoherent metal we find to ordered state(s)),
this local Z$_{2}$ symmetry will control the physical response of the system.
An interesting issue is to inquire about a possible link to fractionalization 
ideas.  Since we have seen 
that non-FL vis-a-vis FL metallicity is caused by an irrelevant (non-FL) 
vis-a-vis relevant (FL) $V_{fc}(k)$, the confinement-deconfinement transition in the $b$-fermion sector is now
tied down to the underlying selective-Mott transition in the $b$-fermion sector, since 
there is no question of having fractionalization in the heavy FL phase.  More 
work is called for to see whether this is correct, or whether the exotic 
excitations represented by the emergent branch-cut form of $\chi_{ii}^{+-}(\omega)$ at the OSMT 
admit no particle description at all~\cite{phillips} - this is  also a known generic 
tendency at QCPs.  However, along with Fermi surface reconstruction (involving
penetration of zeros of the Green function to where poles would have been in the heavy-FL), 
the above arguments hint at an exotic topological criticality 
underlying our findings.  These connections will be explored separately.     
      
\bibliographystyle{apsrev4-1}

\begin{thebibliography}{60}
\bibitem{hertz} J. Hertz, Phys. Rev. B 14, 1165 (1976).
\bibitem{coleman} P. Coleman and A. Schofield, Nature 433,226 (2005).
\bibitem{si} Q. Si et al., Nature 413, 804 (2001).
\bibitem{pepin} C. Pepin, Phys. Rev. B 77, 245129 (2008).
\bibitem{senthil} T. Senthil et al., Phys. Rev. Lett. 90, 216403 (2003).
\bibitem{laad} M. S. Laad et al., Journal of Physics: Condensed Matter 24, 232204 (2012).
\bibitem{civelli} M. Civelli, Ph.D Thesis (Rutgers Univ), arXiv:0710.2802.
\bibitem{georges} S. Biermann et al., Phys Rev Lett. {\bf 95},206401 (2005).
\bibitem{anderson} P. W. Anderson, Nature Physics 2, 626 (2006).
\bibitem{varma} C. M. Varma, Phys. Rev. B 73, 155113 (2006).
\bibitem{liviu} L. Hozoi et al., Phys Rev Lett. {\bf 99}, 256404 (2007).
\bibitem{imada} H. Sakakibara et al., Phys Rev B {\bf 89}, 224505 (2014).
\bibitem{giamarchi} C. Weber et al., Phys Rev Lett. {\bf 112}, 117001 (2014).
\bibitem{cmv} C. M. Varma et al., Phys Rev Lett {\bf 63}, 1996 (1989).
\bibitem{sachdev-ye} N. Read et al., Phys Rev B {\bf 52}, 384 (1995).
\bibitem{alps} B. Bauer et al., Journal of Statistical Mechanics: Theory and Experiment 2011 (2011).
\bibitem{jpcx} S. Acharya {\it et al.}, arXiv: 1602.08990, submitted as a Conf. Proceedings to J. Phys: Conf. Series. 
\bibitem{ong} S. Pathak {\it et al.}, Phys. Rev. Lett. {\bf 102}, 027002 (2009).
\bibitem{schroder} A. Schr\"oder {\it et al.}, Nature 407, 351-355 (21 September 2000).
\bibitem{aeppli} G. Aeppli {\it et al.}, Science, {\bf 278}, 1432-1435 (1997).
\bibitem{pwa} Philip W Anderson and Philip A Casey 2010 J. Phys.: Condens. Matter 22 164201.
\bibitem{akrap} C. C. Homes et al., Nature Scientific Reports 3, Article number: 3446 (2013).
\bibitem{capone} L de Medici {\it et al.}, Phys. Rev. Lett. {\bf 102}, 126401 (2009); S. D. Das {\it et al.},
Phys Rev B {\bf 92}, 155112 (2015) and references therein.
\bibitem{ourFeAs} S. D. Das et al., Phys Rev B {\bf 92}, 155112 (2015).
\bibitem{fisher} T. Senthil and M. P. A. Fisher, Phys. Rev. B 62, 7850 (2000).
\bibitem{casey} Casey, P. A., Anderson, P. W. (2011). Physical Review Letters, 106(9), 097002.
\bibitem{schotte} K. D. Schotte and U. Schotte, Phys. Rev. 182, 479 (1969).
\bibitem{hopfield} J. Hopfield, Comments on Solid State Physics 2, 40 (1969).
\bibitem{mh} E. M\"uller-Hartmann et al., Phys Rev B {\bf 3}, 1102 (1971).
\bibitem{reznik} D, Reznik et al., Nature 440, 1170 (2006).
\bibitem{wang} Y. Wang et al., Phys. Rev. B 73, 025410 (2006).
\bibitem{sudip} S. Chakravarty, Phys. Rev. B 63, 094503 (2001).
\bibitem{tranquada} J. M. Tranquada et al., Nature 375, 561-563 (1995).
\bibitem{fujita} K. Fujita et al., Science 344, 612 (2014).
\bibitem{zaneen} S. Mukhin et al., Phys Rev B  {\bf 76}, 174521 (2007).
\bibitem{khaliullin} G. Khaliullin et al., Physica C282-287, 1751 (1997).
\bibitem{conte} S. Dal Conte et al., Science 335, 1600 (2012).
\bibitem{moon} E. G. Moon and S. Sachdev, Phys. Rev. B 80, 035117 (2009).
\bibitem{keimer} V. Hinkov et al., Science {\bf 319},597 (2008).
\bibitem{kivelson} V. Emery et al., Nature {\bf 374}, 434 (1994).
\bibitem{scalapinolee} P. A. Lee et al., Rev. Mod. Phys. 78, 17 (2006). D. J. Scalapino, Physics Reports 250, 329 (1995).
\bibitem{pint} D. Reznik, Advances in Condensed Matter Physics, vol. 2010, Article ID 523549 (2010).
\bibitem{cDMFT-nematic} M. Tsuchiizu et al., Phys Rev Lett {\bf 111}, 057003 (2013).
\bibitem{gil} S. Sebastian et al., Rep. Prog. Phys. {\bf 75}, 102501 (2012).
\bibitem{kitaev} A. Kitaev, Annals of Physics 303, 2 (2003).
\bibitem{baskaran} G. Baskaran et al., Phys. Rev. Lett. 98, 247201(2007).
\bibitem{moess} J. Knolle et al., Phys Rev Lett {\bf 112}, 207203 (2014).
\bibitem{phillips} K. B. Dave {\it et al.}, Phys. Rev. Lett. {\bf 110}, 090403 (2013).
\end{thebibliography}
 
\end{document}